\documentclass[fleqn,usenatbib]{mnras}

\usepackage{newtxtext,newtxmath}
\usepackage[T1]{fontenc}
\usepackage{ae,aecompl}
\usepackage{multirow,color}
\usepackage{dcolumn}
\usepackage{bm}
\usepackage{amsmath,amsfonts,graphicx,epsf}
\usepackage{color}
\usepackage{enumerate}

\newcommand{\bvec}[1]{\mathbf{#1}}
\newcommand{\pa}{\partial}
\newcommand{\red}[1]{{{#1}}}

\newcommand{\nhour}{4} 

\newcommand{\oneDphiex}{174.8$\pm$10.4~eV}
\newcommand{\strahltestev}{486}
\newcommand{\chisqdof}{1.10}
\newcommand{\dof}{172}
\newcommand{\strahlfail}{5}
\newcommand{\rone}{0.202}

\newcommand{\outlierperc}{0.01} 
\newcommand{\lowenergy}{391} 
\newcommand{\highenergy}{935} 

\title{The Heliospheric Ambipolar Potential Inferred from Sunward-Propagating Halo Electrons}

\author[K. Horaites et al.]{
Konstantinos~Horaites,$^{1}$\thanks{E-mail: konstantinos.horaites@helsinki.fi}
Stanislav~Boldyrev$^{2,3}$
\\
$^{1}$Department of Physics, University of Helsinki, Helsinki, Finland \\  
$^{2}$Department of Physics, University of Wisconsin--Madison, Madison, WI 53706, USA\\
$^{3}$Center for Space Plasma Physics, Space Science Institute, Boulder, CO 80301, USA\\
}

\date{Accepted XXX. Received YYY; in original form ZZZ}

\pubyear{2022}

\begin{document}
\label{firstpage}
\pagerange{\pageref{firstpage}--\pageref{lastpage}}
\maketitle

\begin{abstract}
\red{We provide evidence that the sunward-propagating half of the solar wind electron halo distribution evolves without scattering in the inner heliosphere. We assume the particles conserve their total energy and magnetic moment, and perform a ``Liouville mapping''  on electron pitch angle distributions measured by the Parker Solar Probe SPAN-E instrument.}  
Namely, we show that the distributions are consistent with Liouville's theorem if an appropriate interplanetary potential is chosen.
This potential, an outcome of our fitting method, is compared against the radial profiles of proton bulk flow energy. 
We find that the inferred potential is responsible for nearly 100\% of the proton acceleration in the solar wind at heliocentric distances 0.18-0.79~AU.
These observations combine to form a coherent physical picture: the same interplanetary potential accounts for the acceleration of the solar wind protons as well as the evolution of the electron halo.
In this picture the halo is formed from a sunward-propagating population that originates somewhere in the outer heliosphere by a yet-unknown mechanism.

\end{abstract}


\section{Introduction}

\red{Electrons in the solar wind occupy different collisional regimes, depending on their energy.}  The typical frequency of Coulomb collisions experienced by a test particle is known to fall off precipitously with the particle's speed (as $v^{-3}$). At low energies $\lesssim$10 eV, collisions help to shape the electron velocity distribution function (eVDF) into a Maxwellian ``core''. 
But the so-called ``suprathermal'' electrons, which have speeds much greater than the thermal speed, largely ignore their binary electrostatic interactions with nearby charged particles. It is frequently appropriate to treat the suprathermal electrons as ``collisionless'', so that their motion is only guided by the collective fields in the plasma. 
\red{The influence of these fields on the particles can be loosely separated into two categories: the acceleration by large-scale fields and scattering by wave-particle interactions.} 

The rapid escape of electrons to large heliospheric distances establishes an equilibrium where the near-sun environment has a slight excess of positive charge. \red{The resulting large-scale ``ambipolar'' or ``polarization'' electric field radially accelerates the solar wind ions, which then drag the electrons along.  This field was not considered explicitly in the earliest hydrodynamic models of the solar wind \cite[e.g.,][]{parker58}. But it was gradually incorporated into the theory, where for instance it features prominently in so-called ``exospheric'' or kinetic models \cite[e.g., ][]{sen69, jockers70, lemaire71}.} In these kinetic models, particles abruptly become collisionless above the ``exobase'' \cite[e.g., ][]{zouganelis04, boldyrev20}.  The estimated polarization potential between the exobase ($\sim$0.01-0.05 AU) and 1~AU is on the order of 100-1000~eV.

Despite the general acknowledgment of a large-scale electrostatic field's presence in the solar wind, the magnitude of this field ($E$$\sim$$10^{-9}$~V/m) is far too small to be measured directly by spacecraft. However, recent progress has been made by considering the impact of the electric potential on the eVDF.   
In \citet{bercic21}, abbreviated here as B21, the ambipolar potential was indirectly inferred from eVDFs measured by the Parker Solar Probe (PSP) SPAN-E electron instrument \citep{whittlesey20}. 
\red{These measurements are based on two signals associated with the large-scale potential: 1) the ``deficit'' of sunward-moving electrons in the Maxwellian core \citep{halekas20, halekas21}, and 2) the ``breakpoint energy'' \citep{scudderolbert79a, bakrania20} that delineates the core from the suprathermal electrons.}  In B21, this theory was applied to PSP data to infer a parallel electric field $E_\parallel$ with a typical magnitude $\sim$$10^{-9}$ V/m and scaling with heliocentric distance $E_\parallel\propto r^{-1.69}$. \red{As reported, a solar wind accelerated by such a field would have 59\% of the energy (77\% of the speed) predicted by exospheric models at 45 solar radii ($R_S$). The remaining energy however presents a significant gap between these observations and theory.}

Although the results described above demonstrate a strong connection between the potential and the electron distribution, some limitations of the core deficit method can be recognized. \red{The detailed physics governing the Maxwellian core distribution is complex, as the solar wind electrons are weakly collisional (Knudsen number $\gtrsim$0.01), and exhibit large plasma gradients \cite[e.g., ][]{bale13}. Such a kinetic regime is difficult to treat with standard methods \citep[e.g., ][]{spitzerharm53, gurevichistomin79}. This is before consideration of instabilities that may affect the core \cite[e.g.,][]{schroeder21}}, which may even be generated by a resonant interaction with the deficit electrons themselves \citep{bercic21b}. Thus the authors of B21 caution that their approach is ``simplified and includes strong assumptions'', and may cause either a systematic overestimation or underestimation of the potential depending on how the method is applied. The core deficit in particular is not clearly described theoretically, as this subtle feature is simply identified with the ``electron cutoff'' (a sharp discontinuity) in the sunward eVDF that is seen in exospheric models. Without more detailed predictions, the deficit can only be measured with a heuristic approach. The core perturbation has only been detected in 57\% of eVDFs at heliospheric distances 20--85~$R_S$. This fractional occurrence decreases at the larger radial distances occupied by most spacecraft.

\red{In the current work, we develop a method that complements the pioneering study of B21 by considering the effect of the large-scale potential on a different part of the electron distribution: the suprathermal ``halo'' population.} The halo is identified in solar wind eVDFs as a nearly-isotropic tail at energies $\sim$100-1000 eV \cite[e.g., ][]{feldman75, pilipp87}. \red{As the halo  energies are comparable to predictions for the inner heliospheric potential, this potential should profoundly affect the eVDF.}

In the present work we neglect the processes that generate the halo, but this topic deserves some review. \red{Notably, the apparent growth of the halo at the expense of the anti-sunward suprathermal ``strahl'' population may imply} that the halo is locally formed in the inner heliosphere by scattered strahl electrons \cite[e.g., ][]{maksimovic05, stverak09}. This has led to significant theoretical development, focused on the resonant interaction of electrons with the whistler and fast-magnetosonic whistler (FM/W) modes \citep[e.g., ][]{vocks05, saitogary07, vasko19, verscharen19b, micera21, zentenoquinteros21, vo22, tang22}. 
Observations have struggled to confirm these theories. Notably,  whistlers are practically absent (occurrence rate $<$0.1\%) during PSP perihelion passes \citep{cattell22}. Additionally, the eVDFs sampled by Helios and PSP are stable with respect to the oblique FM/W mode \citep{jeong22}. Theoretical calculations show that at $r\lesssim$1 AU the strahl is stable to whistler fluctuations \citep{horaites18b, schroeder21} and should be unaffected by whistler turbulence in the inner heliosphere \citep{boldyrev19}. High-resolution measurements of the strahl at 1~AU confirm that ``anomalous diffusion'', e.g. from whistler waves, is not required to explain the strahl angular widths at resolvable energies $\lesssim$300 eV \citet{horaites18, horaites19}. Similar results were found from simulations at distances  $r\lesssim$20$R_S$ \citep{jeong22b}, which showed that the strahl is adequately described by Coulomb collisions near the corona. This all suggests that a mechanism besides local wave particle scattering may account for the halo's presence in the inner heliosphere. Such theories have been proposed \cite[e.g., ][]{leubner05,lichko17,horaites19, che19,scudder19}, though no consensus has emerged.

Leaving the halo's precise origin aside, we here adopt a very simple model. We assume that the sunward-moving halo electrons originated from some heliocentric distance beyond the view of PSP ($>$0.8 AU), as it is natural for highly energetic sunward-moving particles to originate from larger distances. 
\red{From the considerations detailed above, we are motivated to neglect scattering effects. The halo electrons will then be treated in the context of steady-state collisionless theory. }

Neglecting both Coulomb collisions and wave-particle interactions, we will apply Liouville's theorem, which states that the eVDF is conserved along the electron trajectories. The electrostatic field essentially causes a shift in the halo energy spectra, and this energy shift directly corresponds with the ambipolar potential itself. We will therefore perform a ``Liouville mapping'' \cite[e.g., ][]{schwartz98, lefebvre07} to infer the potential from the eVDFs as observed by Parker Solar Probe in the inner heliosphere 0.18--0.79 AU (39--170 $R_S$).

We will demonstrate that a single Liouville mapping can be applied to accurately model the halo eVDFs over a broad range of distances.  Moreover, we will show that the ambipolar potential inferred from the eVDFs is exactly that required to accelerate the solar wind protons over these distances. This creates a coherent physical picture in which the same electric field causes two independently measurable trends as heliocentric distance increases: the halo loses energy  while the bulk proton flow accelerates.

\section{Theory}\label{theory_sec}

\red{The Vlasov equation describes the evolution of the distribution function $f(\bvec{x}, \bvec{v}, t)$ for a particle species in the absence of diffusion:}

\begin{equation}\label{vlasov_eq}
\dfrac{\pa f}{\pa t} + \bvec{v}\cdot \nabla_x f + \bvec{a} \cdot \nabla_v f = 0,
\end{equation}

\noindent where $\bvec{a}(\bvec{x}, t)$ represents the acceleration due to the forces (Lorentz, gravitational) that act on the particles. \red{Equation~\ref{vlasov_eq} neglects both Coulomb collisions and wave-particle interactions. The general solution, known also as Liouville's Theorem, states that $f$ is constant along the particle trajectories:}

\begin{equation}\label{liouville_theorem_eq}
f(\bvec{x}(t), \bvec{v}(t), t) = f(\bvec{x}_0, \bvec{v}_0, t_0),
\end{equation}

\noindent In eq.~\ref{liouville_theorem_eq}, the coordinates $\bvec{x}(t)$, $\bvec{v}(t)$, $t$ describes the time-dependent motion of an arbitrary particle with initial position $\bvec{x}_0 = \bvec{x}(t_0)$ and velocity $\bvec{v}_0 = \bvec{v}(t_0)$, under the acceleration $d\bvec{v}/dt =\bvec{a}(\bvec{x}(t), t)$. 

Liouville's Theorem~(\ref{liouville_theorem_eq}) is especially useful when the particle trajectories $\bvec{x}(t)$, $\bvec{v}(t)$ observe some constants of motion ($c_1$, $c_2$, ...). In such a case it is possible to recast $f$ in terms of those constants \cite[e.g., ][]{pierrardlemaire96}, so that the function $f=\tilde f(c1, c2, ...)$ is a solution to the steady-state ($\pa f / \pa t$=0) Vlasov equation~(\ref{vlasov_eq}). 

Let us consider a steady-state solar wind in the ecliptic, where the magnetic field $B$, ambipolar potential $\phi$, and the halo electron distribution $f$ vary spatially with the heliocentric distance $r$. For halo electrons in the solar wind, the relevant constants of motion are the total energy $\mathcal{E}$ and magnetic moment $M$:

\begin{equation}\label{E_eq}
\mathcal{E} = K - e \phi(r),
\end{equation}

\begin{equation}\label{M_eq}
M = \frac{K \sin^2 \theta}{B(r)},
\end{equation}

\noindent In eqs.~\ref{E_eq}-\ref{M_eq}, variables $K$ and $\theta$ represent the electron kinetic energy ($K=m_ev^2/2$) and pitch angle with respect to the magnetic field, respectively.  We also introduce $e$, the modulus of the elementary charge.
Note that in eq.~\ref{E_eq} the Sun's gravitational potential is ignored, as the gravitational energy of an electron at 1~AU is $\ll$1~eV.

\begin{figure}
\includegraphics[width=1\linewidth]{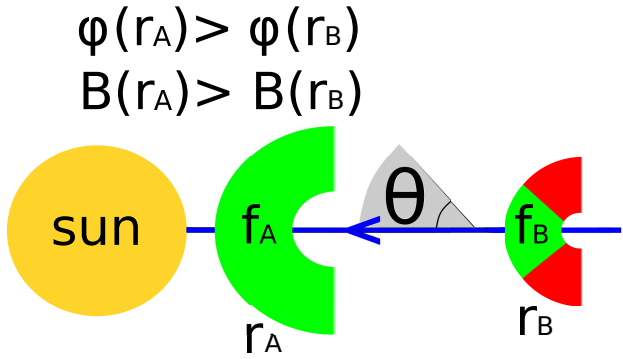}
\caption{\label{liouville_fig} 
As a consequence of Liouville's theorem, an initial halo electron distribution at position $r_B$ will distort as the particles migrate towards the Sun along a field line (blue). This process is shown schematically. The green wedge-shaped region of the velocity distribution $f_B$ (imagined in $v_\parallel$,$v_\perp$ space) maps to the distribution $f_A$. The mirror force causes broadening, while the changing potential causes the electrons to gain energy. The red regions of $f_B$ represent particles that will be reflected by the magnetic field before arriving at $r_A$. The white semicircular vacancies in $f_A$,$f_B$ are the domain of the collisional core distribution, not considered here. }
\end{figure}

In the solar wind we expect the large scale gradients $d B/dr<$0 and $d\phi/dr<$0. Consequently, as a halo electron travels towards the Sun its pitch angle and kinetic energy will both tend to increase as the particle attempts to conserve its values of $\mathcal{E}$ and $M$. The phase space density is conserved along the electron trajectories, so the sunward-moving half of the (gyrotropic) distribution function evolves with distance as shown in Fig.~\ref{liouville_fig}.

Note that the 2D velocity space of the distribution $f$ may be written in terms of many sets of variables, e.g., $\{v_\perp$,~$v_\parallel\}$, ${\{v,~\cos\theta\}}$, or ${\{K,~\theta\}}$. In the context of Liouville's theorem it is often convenient to use $\{\mathcal{E}, M\}$. For the purpose of comparison with spacecraft measurements, in this work we will primarily use $\{K, \theta\}$.

As discussed above, conservation of $\mathcal{E}$, $M$ leads to a general solution to the steady-state ($\pa f/\pa t$=0) Vlasov equation (\ref{vlasov_eq}) for the halo electrons:

\begin{equation}\label{f_E_M_gen_eq}
f=\tilde f(\mathcal{E}, M).
\end{equation}

\noindent Given such a function $\tilde f$, the solution $f(r, K, \theta)$ can be written immediately by substituting the definitions~(\ref{E_eq}), (\ref{M_eq}) into (\ref{f_E_M_gen_eq}):

\begin{equation}\label{f_E_M_eq}
f(r,K,\theta) = \tilde f\Big(\hspace{0.1cm} K - e \phi(r),\hspace{0.2cm}\frac{K \sin^2 \theta}{ B(r)} \hspace{0.1cm}\Big).
\end{equation}

\noindent In terms of the variables $r$, $K$, $\theta$, Liouville's theorem requires that any solution for $f$ should be of the form (\ref{f_E_M_eq}), which just restates that $f$ is a function of the total energy and the magnetic moment. For simplicity we will only consider sunward-directed electrons, i.e. $\theta\in$~[0$^\circ$, 90$^\circ$]. This avoids accounting for the phase-space redundancy $\sin^2\theta = \sin^2(180^\circ - \theta)$, \red{which would complicate the notation. } We also treat the ``passing'' and ``reflected'' populations \cite[e.g., ][]{lefebvre07} equally, as the distinction is not of great importance here.

Frequently, one wishes to map the eVDF from one location to another. Let us suppose that at a given position $r_\star$ the functional form of the eVDF, $f^\star(K, \theta)$, is known:

\begin{equation}\label{f_star_def_eq}
f^\star (K, \theta) \equiv f(r_\star,K,\theta).
\end{equation}

\noindent \red{Then, the solution $f(r,K,\theta)$ is constructed by repeatedly substituting the definitions~(\ref{E_eq}), (\ref{M_eq}) of the invariants into the boundary condition~(\ref{f_star_def_eq}):}



\begin{equation}\label{liouville_map_eq}
\begin{split}
f(r,&K,\theta) = \\
& f^\star\bigg( \hspace{0.1cm}K - e\Delta\phi(r,r_\star)\hspace{0.1cm}\text{{\huge ,}}\hspace{0.2cm}\sin^{-1} \sqrt{\frac{ B(r_\star)K \sin^2 \theta}{B(r)(K - e\Delta\phi(r,r_\star))}}\hspace{0.1cm} \bigg),
\end{split}
\end{equation}

\noindent where we have introduced the notation $\Delta\phi(r_A, r_B)$ to represent the potential difference between distances $r_A$ and $r_B$,

\begin{equation}\label{delta_phi_def_eq}
    \Delta \phi(r_A, r_B)\equiv \phi(r_A)-\phi(r_B).
\end{equation}

\noindent It can be verified by inspection that the mapping formula~(\ref{liouville_map_eq}) has the form~(\ref{f_E_M_eq}) required by Liouville's theorem and satisfies the boundary condition~(\ref{f_star_def_eq}). The arguments of the 2D function $f^\star$ in (\ref{liouville_map_eq}) are the formulas for how K, $\theta$ of a given electron respectively map across distance under the constraints $\mathcal{E}$=const., $M=$const. Technically, our Liouville mapping is derived for eVDFs measured simultaneously along the same field line, but we ignore this detail by assuming the solar wind in the ecliptic is steady-state and cylindrically symmetric.

From equations~\ref{f_E_M_eq},~\ref{liouville_map_eq} it is readily seen that if the halo eVDFs are measured at two different distances along with the local magnetic field $B$, the only unknown variable is the potential~$\phi$. With an accurate determination of $\phi$, a function of the form (\ref{f_E_M_eq}) can be found that matches two (or more) such eVDFs.  Such a ``Liouville mapping'' may be accomplished with statistical methods (see sections~\ref{1d_sec},  \ref{2d_sec}) and constitutes a measurement of the potential. \red{In section~\ref{obs_sec} we apply a Liouville mapping to the average eVDFs measured by PSP, to infer the potential  $\phi(r)$ in the inner heliosphere.}

\section{Observations}\label{obs_sec}
Our primary data set comes from the SPAN-E electron experiment, a pair of electrostatic analyzers (ESAs) onboard the PSP satellite \citep{whittlesey20}.  SPAN-E is part of the Solar Wind Electrons Alphas and Protons (SWEAP) investigation \citep{kasper16}.
Each SPAN-E ESA has a nominal field of view (FOV) of 240$^\circ$$\times$120$^\circ$, but the instruments are arranged in a complementary fashion so that together they observe $>$90\% of the sky after considering physical obstructions such as the spacecraft heat shield.
We studied the Level~3 pitch angle distribution (PAD) data, which were generated by collating the 3D eVDFs measured by the two ESAs into 2D ($K,~\theta$) distributions. The pitch angle $\theta$ is determined in the Level~3 data by referencing the FIELDS fluxgate magnetometer \citep{bale16}. Each PAD has 12 evenly spaced pitch-angle bins (15$^\circ$ wide each) and 32 energy bins logarithmically spaced over a range ~$\sim$2--1800 eV. Such PADs are appropriate for the study of solar wind halo, as the halo electrons are gyrotropic and have typical energies 100-1000 eV.

We considered available Level 3 SPAN-E observations ($\sim$$10-20$~second cadence) between the dates October 31, 2018 and December 31, 2020. This time range is centered approximately around the solar minimum associated with the onset of Solar Cycle 25. During this time period the PADs are only available irregularly, but as the spacecraft completed multiple orbits there is sufficient coverage of the distances 0.18-0.79 AU (i.e. each distance is sampled by a few orbits). We remove data associated with co-rotating interaction regions (CIRs) and stream interaction regions (SIRs) encountered by PSP, as indexed by \citet{allen21}. Coronal mass ejections (CMEs) encountered by PSP are also removed using the HELIO4CAST ICMECAT list \citep{moestl20}. Guided by the criteria in \citet{nieves18}, it contains only ICME events that show clear signatures of magnetic obstacles. This helps for instance to avoid contamination by the ``counterstreaming'' strahl, which is a sunward-directed electron beam that is frequently seen during the passage of CMEs \citep{gosling87, gosling92}. Measurements taken while SPAN-E's mechanical attenuator was deployed were also excluded from our data set, because the action of the attenuator reduces the ESA signal enough that the halo distribution can become contaminated with noise \citep{whittlesey20}. As the attenuator was usually deployed during PSP's perihelia passes, it is infeasible to investigate the halo at distances $r\lesssim$0.18 AU.

In this study we compare electron data with proton data measured by SWEAP's Solar Probe Cup (SPC) \citep{case19}. Therefore we require SPC data (i.e., the COHO hourly data product) to be available within~30 minutes of each SPAN-E PAD. \red{We make a rough correction to isolate the slow wind, which has a different origin and composition than the ``fast wind'' \cite[e.g., ][]{verscharen19}. After removal of transient events, we excluded any data for which the radial solar wind (proton) velocity exceeded the 95th percentile at that distance \cite[e.g.,][]{mcgregor11, larrodera20}---this mitigated the fast wind peak in the ``bimodal'' distribution of proton bulk speeds.} 

As we are interested in finding the variation of the potential over large scales via a Liouville mapping, we compute the average PAD at different distances. We bin the PADs into N=8 logarithmically spaced distances 0.18-0.79 AU to compute the average PAD for each bin. Each average PAD is associated with a nominal distance $r_k$, which are indexed sequentially ($r_1$<$r_2$<...<$r_N$). The PADs, which are supplied by SPAN-E in terms of differential energy flux, are converted into phase space density (units m$^{-6}$s$^3$) and are removed of outliers (\outlierperc\% of data) before averaging. 
\red{ We do not correct for the floating spacecraft potential as this should have a negligible effect ($\sim$2 eV) on the measured energies \citep{bale20}. 

Before averaging, we re-orient the Level~3 PADs according to the 1-minute FIELDS magnetometer data so that $\theta$=0$^\circ$ corresponds with sunward-moving field-aligned electrons. We then test for the strahl by checking that the phase space density in the \strahltestev~eV channel is greater for the strahl direction ($180^\circ$) than for the sunward direction ($0^{\circ}$)---any data that do not meet this test are removed ($<$\strahlfail\%).} This process separates the anti-sunward strahl population ($\theta\approx 180^\circ$) from our measurements ($\theta<90^\circ$) of the sunward-oriented halo. The occasional ``sunward strahls'' associated with magnetic switchbacks \citet{macneil20} are consequently removed as well.  In the analysis presented here, we consider only energies for which the phase space density at $\theta$$\approx$0$^\circ$ was in the range [2e-20,1e-17]~m$^{-6}$s$^3$.  This yields spectra with energies 200-1000~eV, avoiding the low-energy core as well as the high-energy ``superhalo'' \citep{lin98, wang12}.

Technically, we pre-averaged the PADs into \nhour-hour long intervals before computing the average over distances. This helped to reduce bias that might arise due to uneven time cadence, and smoothed over short small-scale variability \cite[e.g., ][]{gazis99}. We chose \nhour~hours as a pre-averaging interval because this corresponds with the eddy turnover time in the slow solar wind \citep{weygand13}, i.e. the time-scale over which successive measurements become uncorrelated.
\red{We assume that \nhour-hour averages of solar wind data are independent and identically distributed (i.i.d.). This allows error bars of the distance-averaged PADs to be computed from the standard deviation of the mean. We expect i.i.d. is a coarse approximation as PSP measurements are affected by many processes on different timescales, e.g. turbulence (hours or less), data gaps (days), and solar rotation (weeks). The solar cycle, which occurs over an 11-year period, is unlikely to affect our analysis since our data are all measured near solar minimum. }

\begin{figure}
\includegraphics[width=1\linewidth]{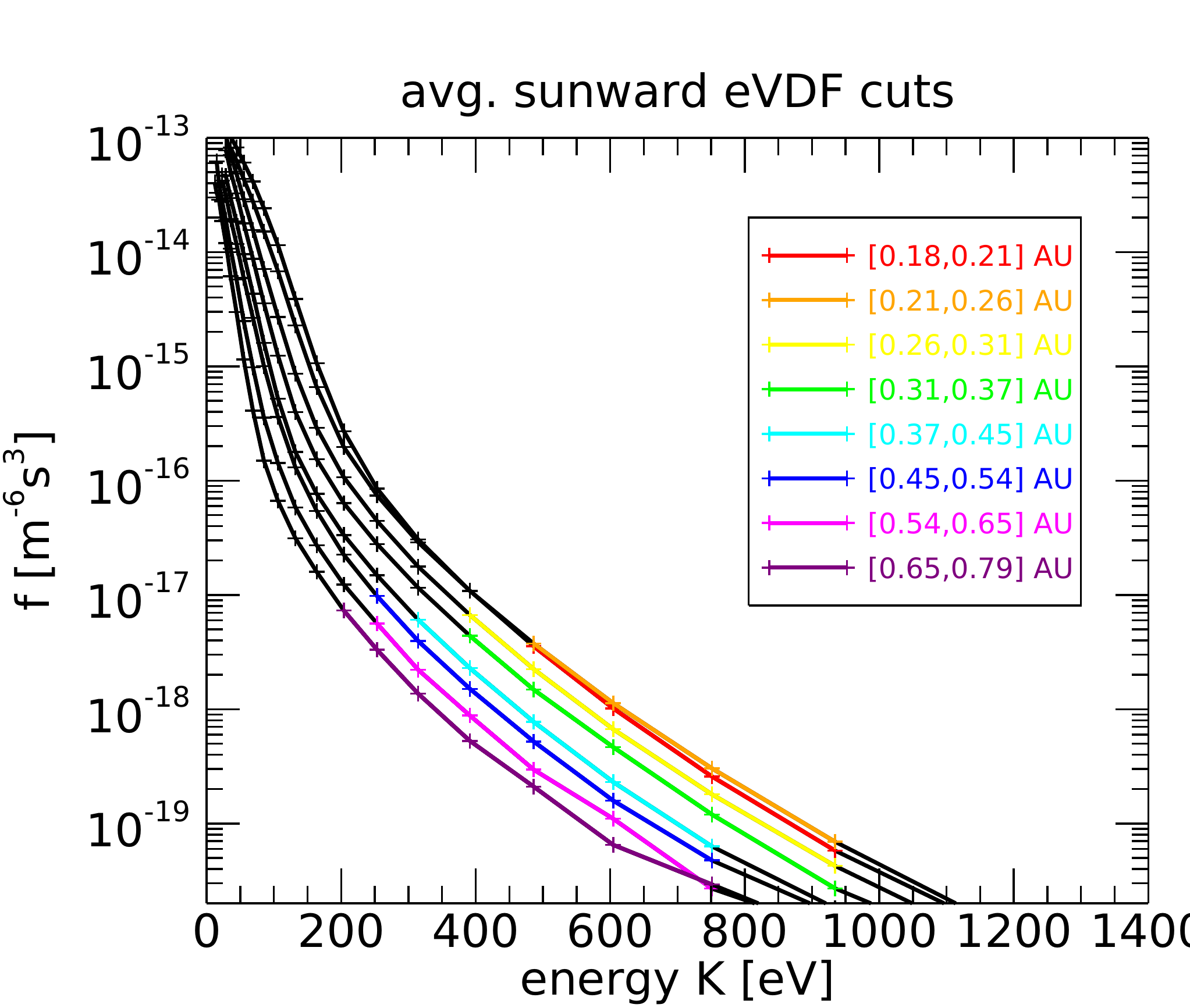}\\
\includegraphics[width=1\linewidth]{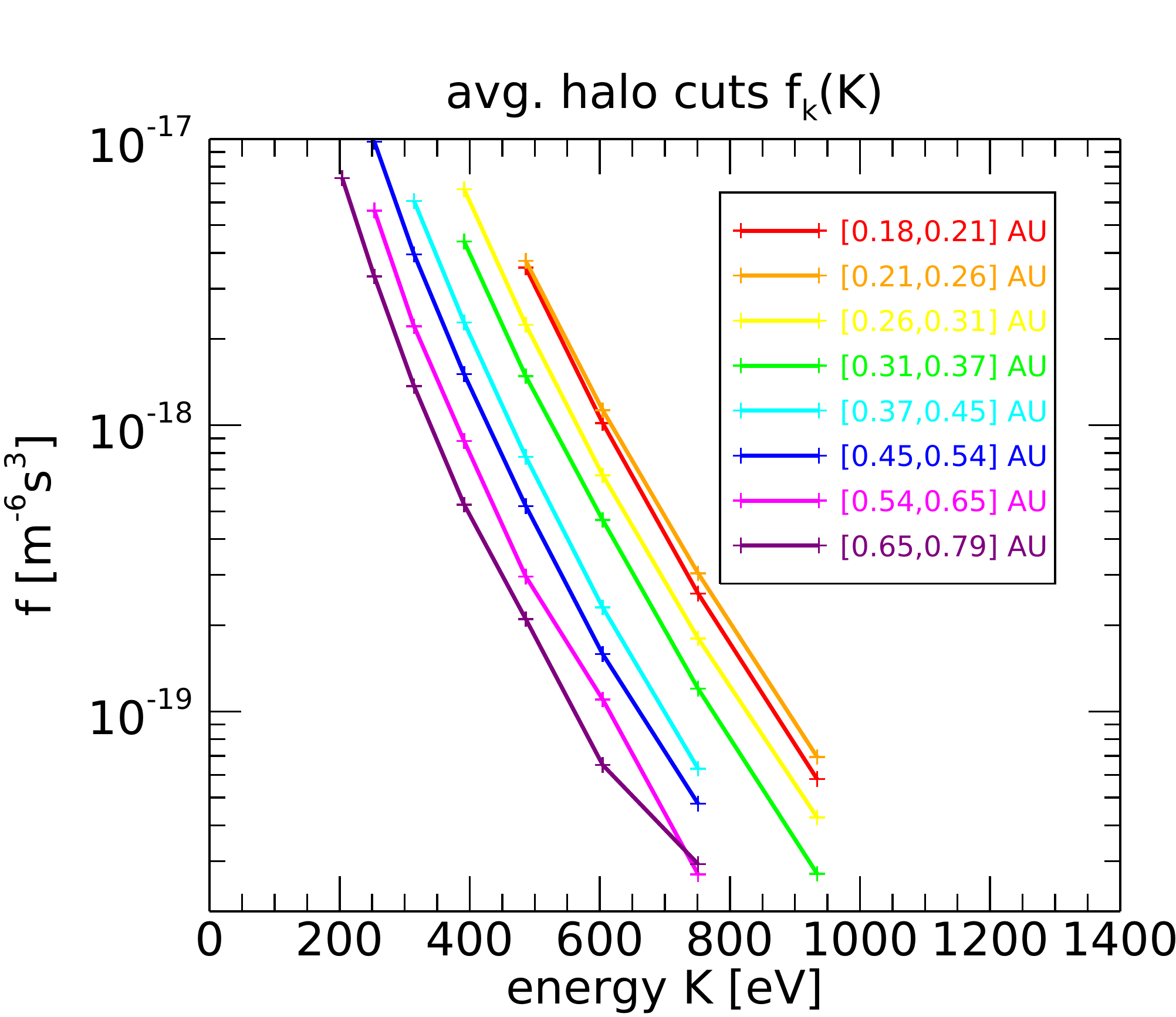}
\caption{\label{f_ave_fig} 
\red{{\bf Top:} Sunward-directed field-aligned eVDF cuts (pitch angle bin $\theta$$\in$$[0^\circ,15^\circ]$). The different lines describe the average PAD measured at different distances ($r_1$, $r_2$, ... $r_N$=$r_8$), which are labeled by color. The legend shows the range of $r$ averaged in each distance bin. The colored sections of the plot indicate the halo regime used for the 1D fitting analysis (section~\ref{1d_sec}). The continuation of the PAD averages outside our domain of interest (black) are not considered in our analysis. {\bf Bottom:} 
The same averaged cuts, focusing only on the halo regime. It can be observed that the halo has a similar shape at each distance, up to an energy shift.}}
\end{figure}

Energy cuts of the resulting distance-averaged PADs are shown in Figure~\ref{f_ave_fig}, for different distances. The energy cuts of the distribution are taken from the most field-aligned sunward pitch angle bin of the PAD (0$^\circ$<$\theta$<15$^\circ$). 
It can be seen that at any given energy the phase space density of the halo decreases with the heliocentric distance $r$. Conversely, as it is relevant for our analysis, we note that for a fixed phase space density the energy decreases with distance. This is just the expected signature given by Liouville's theorem for electrons moving in a potential gradient $d\phi/dr<0$.

We performed a Liouville mapping using two different methods, which we will call ``1D'' and ``2D''. These methods yield similar estimates of the potential $\phi(r)$, but the 1D technique is more illustrative and simple to implement while the 2D method is more general as it accounts for the angular variation of the eVDF.

\subsection{1D Liouville Mapping}\label{1d_sec}

A changing potential causes a corresponding shift in the eVDF. In the 1D Liouville mapping, we consider only the electrons that propagate towards the Sun with pitch angles nearly aligned with the magnetic field (i.e., $\theta \approx$0$^\circ$). We may approximate that the kinetic energies of such electrons change, but their pitch angles stay field-aligned.  It may be readily shown that as a consequence of Liouville's theorem~(\ref{f_E_M_eq}), the 1D energy cuts $f_A(K)=f(r_A,K,0^\circ)$ and $f_B(K)=f(r_B,K, 0^\circ)$  should look identical up to an energy shift due to the potential difference $\Delta\phi$:

\begin{equation}\label{1d_map_eq}
    f_A(K) = f_B(K - e\Delta\phi(r_A,r_B)),
\end{equation}

\noindent so the shift between two energy spectra gives the potential $\Delta \phi$.




We apply the technique described in \citet{horaites21} to measure the energy shift between two 1D eVDF cuts.  Given two discrete 1D energy spectra $f_A(K)$, $f_B(K)$ measured by SPAN-E at respective distances $r_A$, $r_B$, the following procedure is applied:

\begin{enumerate}
 \item A linear piecewise interpolation is performed on the spectrum $f_B$ to yield $f_{int}$. The interpolation energies $K_{int, j}$ are chosen so that each of the N data points in $f_{int}$ has a matching point in $f_A$ with the same phase space density, at energy $K_{A,j}$.
 \item For each such matched pair, the energy difference ${\Delta K_j = K_{A,j} - K_{int,j}}$ is computed. The uncertainty of this difference, $\sigma_{\Delta K, j}$, is computed by propagating errors through the interpolation calculation. This $\sigma_{\Delta K, j}$ depends on the uncertainties in $f_A$ and $f_B$, the sampled energies of the spectra, and the intrinsic energy uncertainty of the SPAN-E detector (7\%).
 \item The ``weighted mean'' \citep[e.g., ][]{barlow89} of the energy difference $\Delta K_j$ is computed, which in the context of our Liouville mapping (\ref{1d_map_eq}) is simply the energy shift $e\Delta \phi$:
\end{enumerate}

\begin{equation}\label{phi_eq}
e\Delta\phi(r_A, r_B) = \frac{ \sum_{j=1}^{N} ( \Delta K_j / \sigma^2_{\Delta K,j})}{  \sum_{j=1
}^{N} (1 / \sigma^{2}_{\Delta K,j} )}.
\end{equation}

\begin{figure}
\begin{centering}
\includegraphics[width=.7\linewidth]{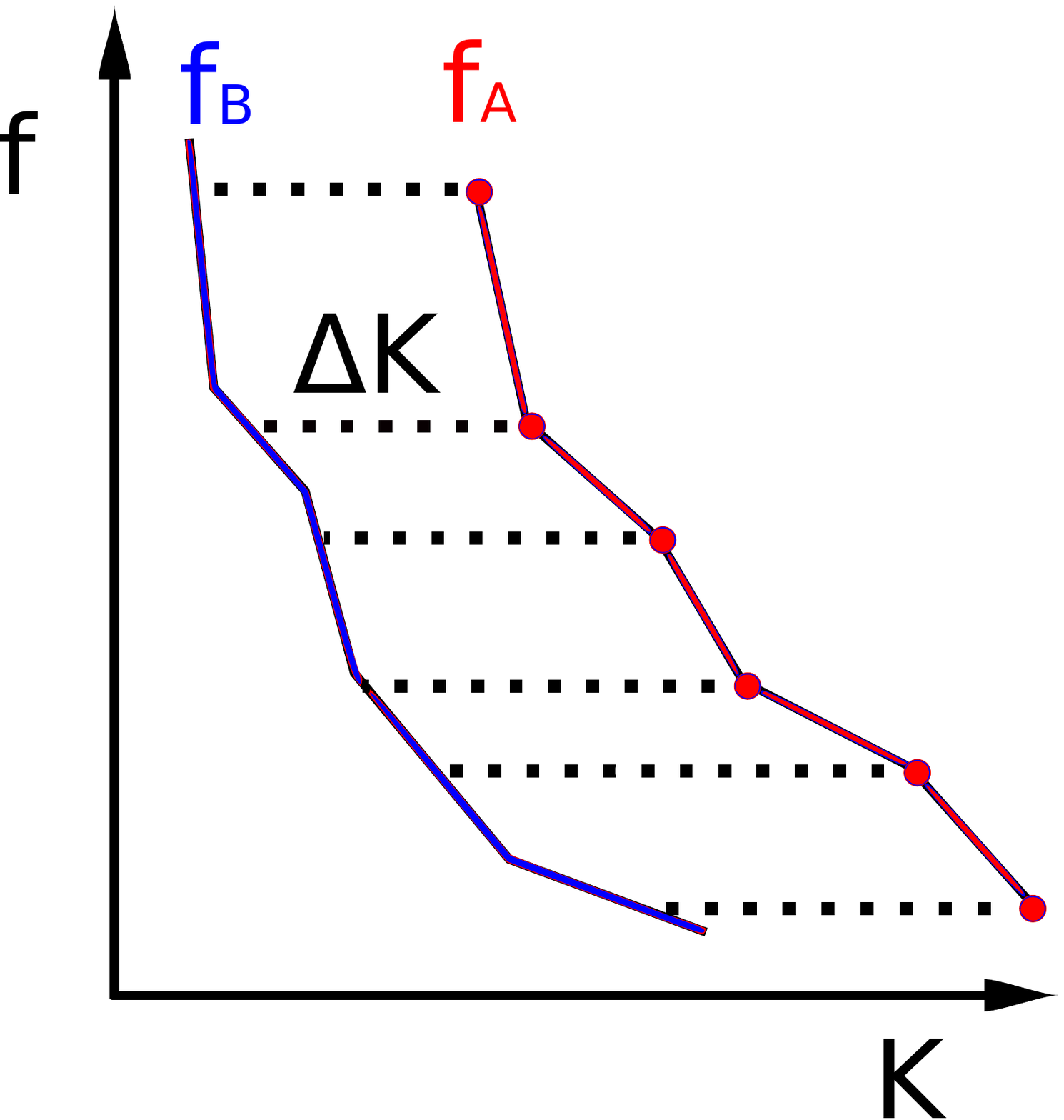}\\
\includegraphics[width=1\linewidth]{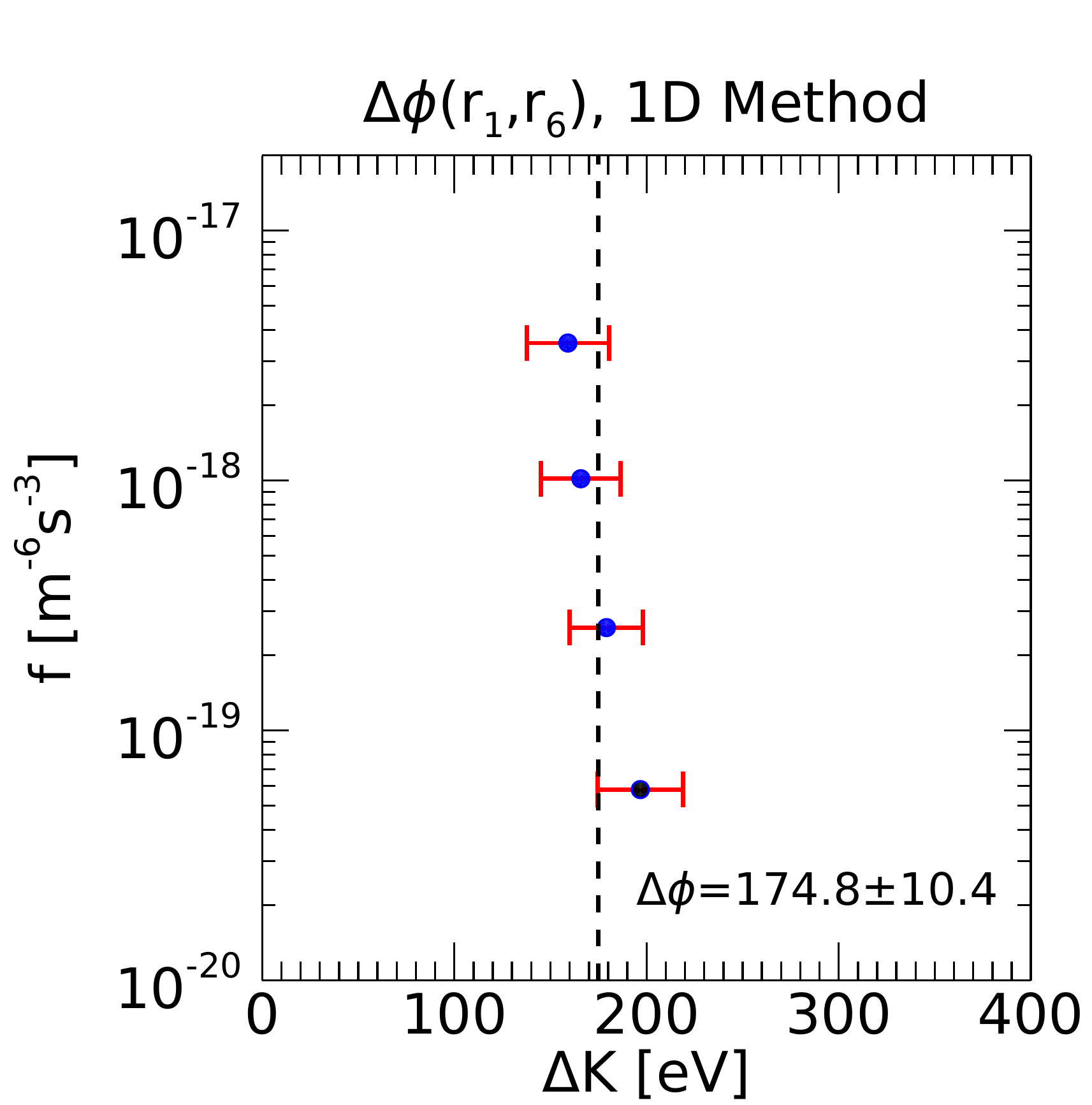}
\end{centering}
\caption{\label{1d_fit_fig} 
{\bf Top:} In the 1D method (section~\ref{1d_sec}), interpolation between two sunward-directed $\theta$$\approx$$0^\circ$ halo cuts at fixed phase space density (y-axis) yields a set of measurements $\Delta K_j$ of the energy shift (dashed lines). These are used to estimate the potential difference $\Delta\phi(r_A, r_B)$ between the distances $r_A$, $r_B$.
{\bf Bottom:}
An example of the 1D method applied to our PAD data, showing measurements of the energy shift $\Delta K_j$ (blue dots) between the $r_1$$\approx$0.2~AU and $r_6$$\approx$0.5~AU cuts. A weighted average of the $\Delta K_j$ (eq.~\ref{phi_eq}) yields the potential difference $\Delta\phi(r_1, r_6)$=\oneDphiex. The final results for the 1D method potentials, $\Delta \phi(r_1,r_k)$ for each $r_k$, are shown in Table~\ref{plasma_param_and_phi_table}.}
\end{figure}

Figure~\ref{1d_fit_fig} illustrates the 1D Liouville mapping described above, for a particular case. The data in the lower panel are derived from the red ($r_1\approx$0.2 AU) and dark blue ($r_6\approx$0.5 AU) curves of Fig.~\ref{f_ave_fig}. Taking these two energy spectra as $f_A$, $f_B$ respectively, we show the energy differences $\Delta K_j$ between $f_B$ and the interpolated spectrum $f_{int}$, in Figure~\ref{1d_fit_fig}. 

\red{We apply the 1D Liouville mapping to the distance-averaged eVDFs to compute the change in potential $\Delta \phi$ (eq.~\ref{phi_eq}). We use the most field-aligned cut of the eVDF  (representing $0^\circ$$<\theta<$15$^\circ$).  The eVDF at the sunward-most distance (at $r_A=r_1$$\approx$0.2 AU) is used as the reference ($f_A$). This is compared to the corresponding $\theta\approx0^\circ$ cut ($f_B$) for each other distance $r_2$,$r_3$,...$r_N$, giving an estimate of $e\Delta \phi$ at each distance. The results are shown in Table~\ref{plasma_param_and_phi_table}. We find that the potential decreases by $\sim$250~V from 0.18 to 0.79 AU.

It should be noted that the formula for the energy shift, eq.~\ref{1d_map_eq}, only works for exactly field-aligned particles ($\theta=0^\circ$). Since an ESA can only sample a finite range of pitch angles, the 1D method has some systematic error because pitch angle broadening from moment conservation is not accounted for. This error manifested as a negative correlation between $f$ and $\Delta K$ (somewhat visible in Fig.~\ref{1d_fit_fig}). This error was generally small ($\sim$10\%), because the average PADs deviate only slightly from isotropy.
The 2D method, described below, is resilient against this error.
}

\subsection{2D Liouville Mapping}\label{2d_sec}

The 2D mapping is performed by fitting a function $f(r, K, \theta)$ of the form~(\ref{f_E_M_eq}) to the full (0$^\circ$<$\theta$<90$^\circ$) sunward PADs. This single fit function describes the eVDF at all distances observed. We will assume that at the outermost distance $r_N$ the logarithm of the distribution $f(r_N, K, \theta)$ can be matched to a 2D polynomial:



\begin{equation}\label{f_poly_eq}
    \ln f^\star(K, \theta) = \ln f(r_N, K, \theta) = \sum_{i=0}^D \sum_{j=0}^D A_{ij} \theta^i K^j.
\end{equation}

\begin{figure*}
    \includegraphics[width=.35\textwidth]{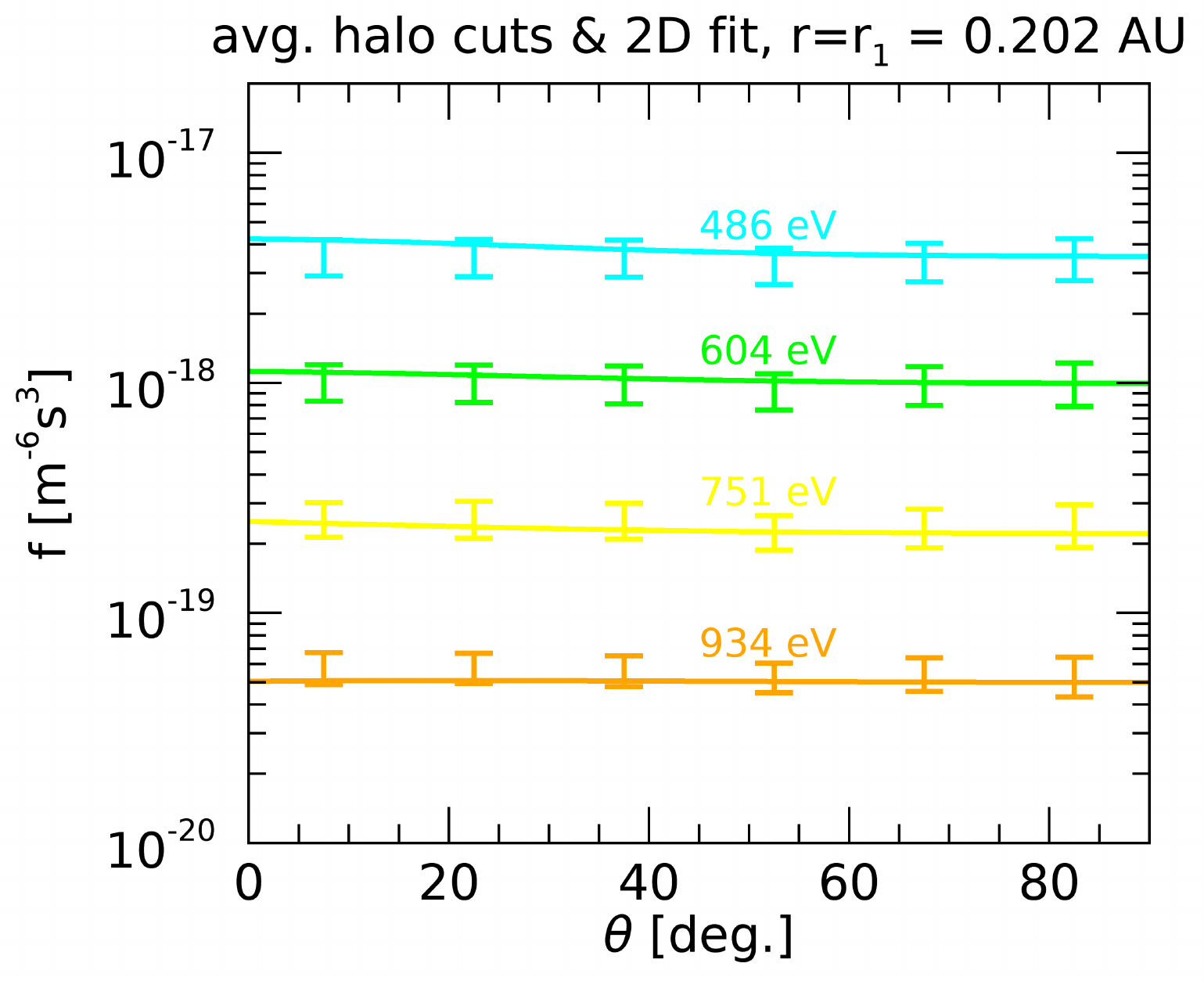}\hspace{1cm}
    \includegraphics[width=.35\textwidth]{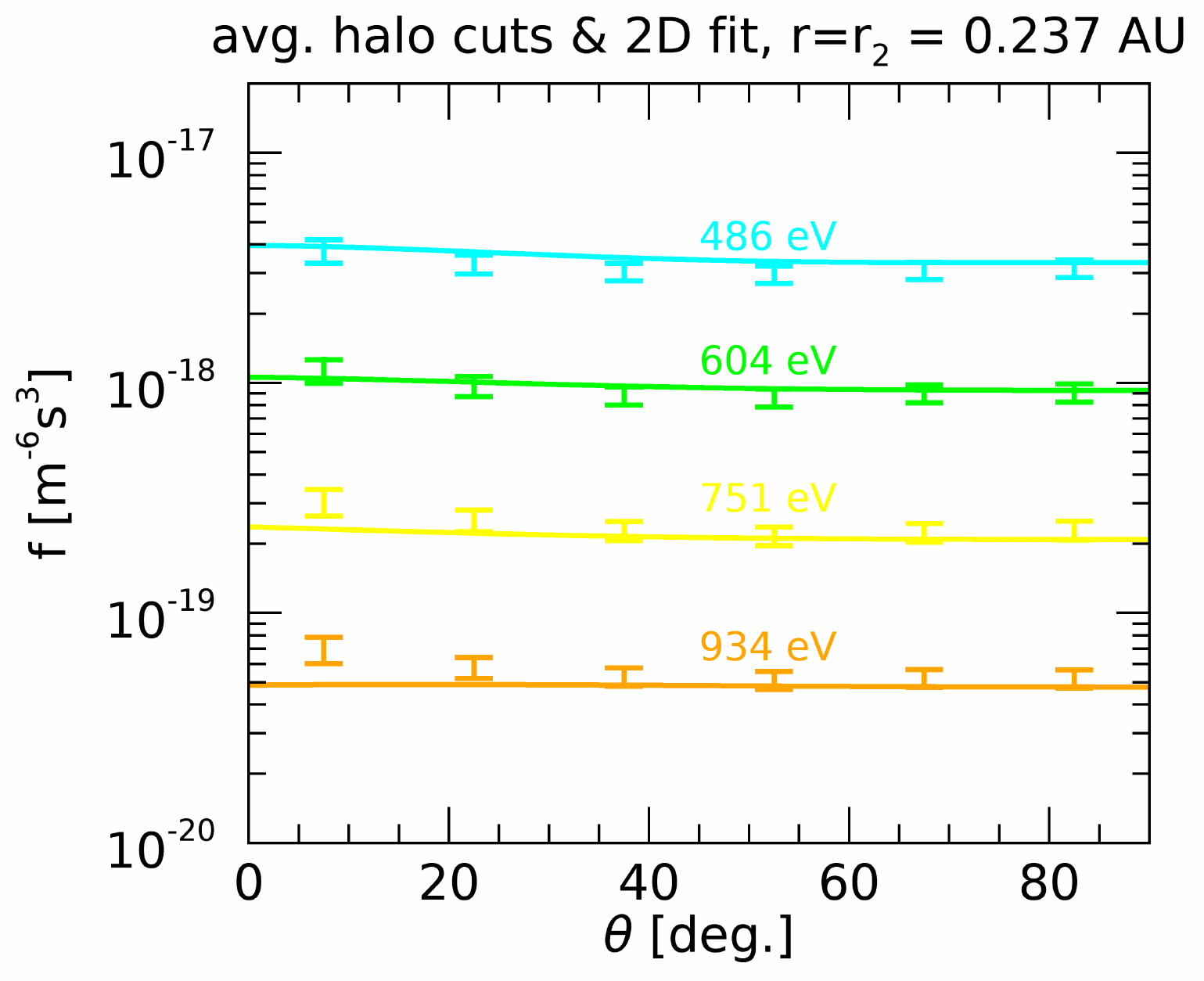}\hfill
    \\[\smallskipamount]
    \includegraphics[width=.35\textwidth]{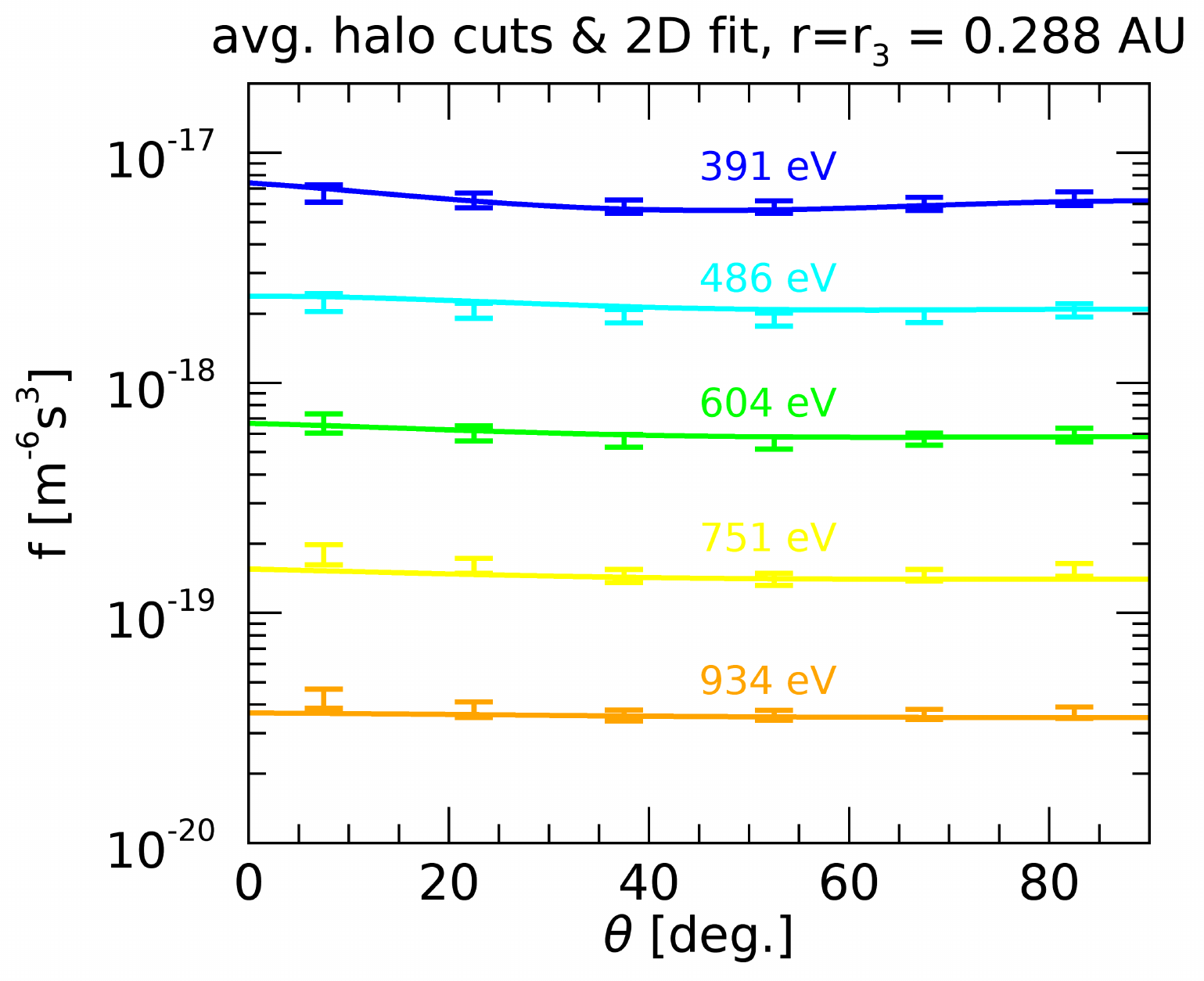}\hspace{1cm}
    \includegraphics[width=.35\textwidth]{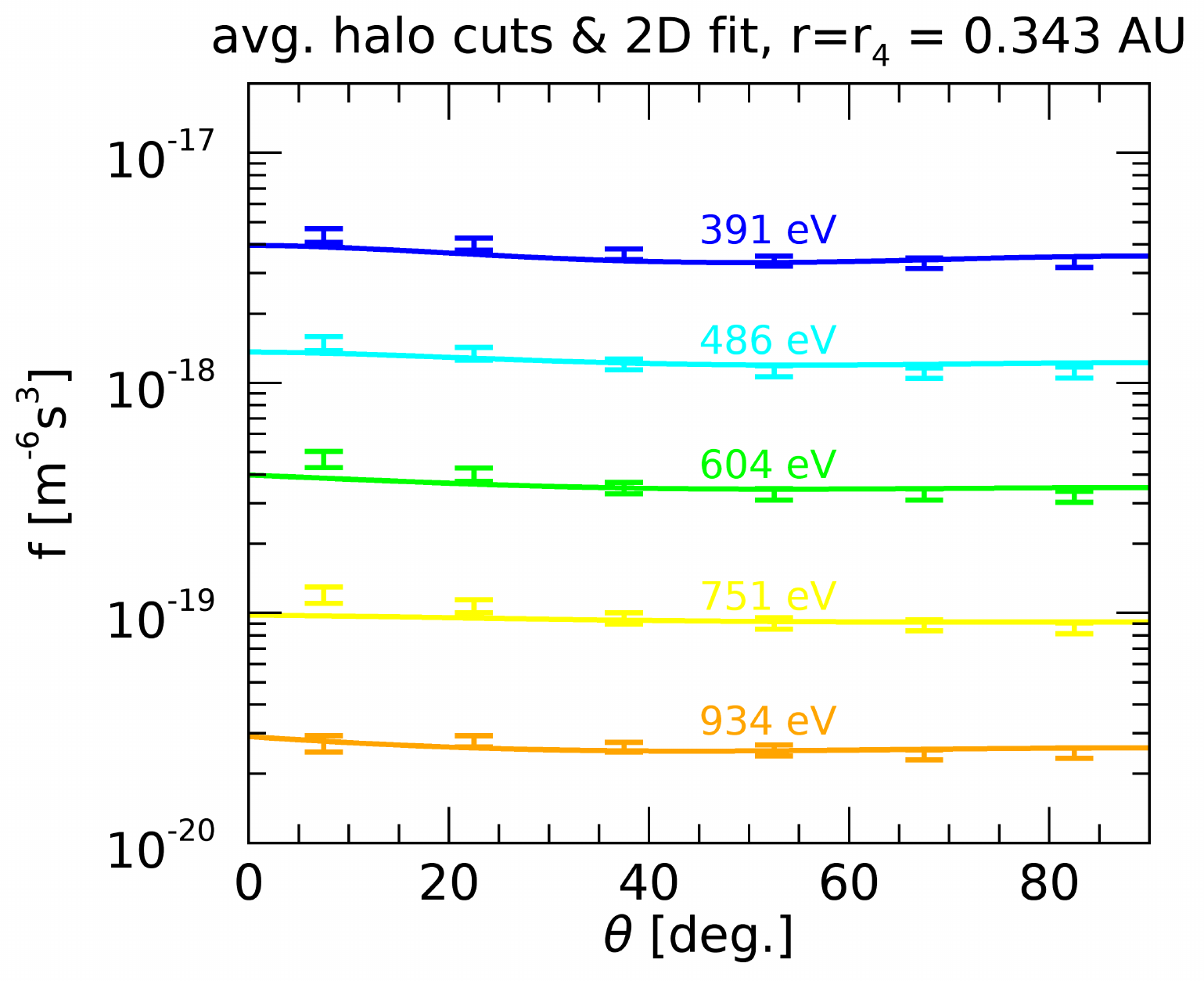}
    \\[\smallskipamount]
    \includegraphics[width=.35\textwidth]{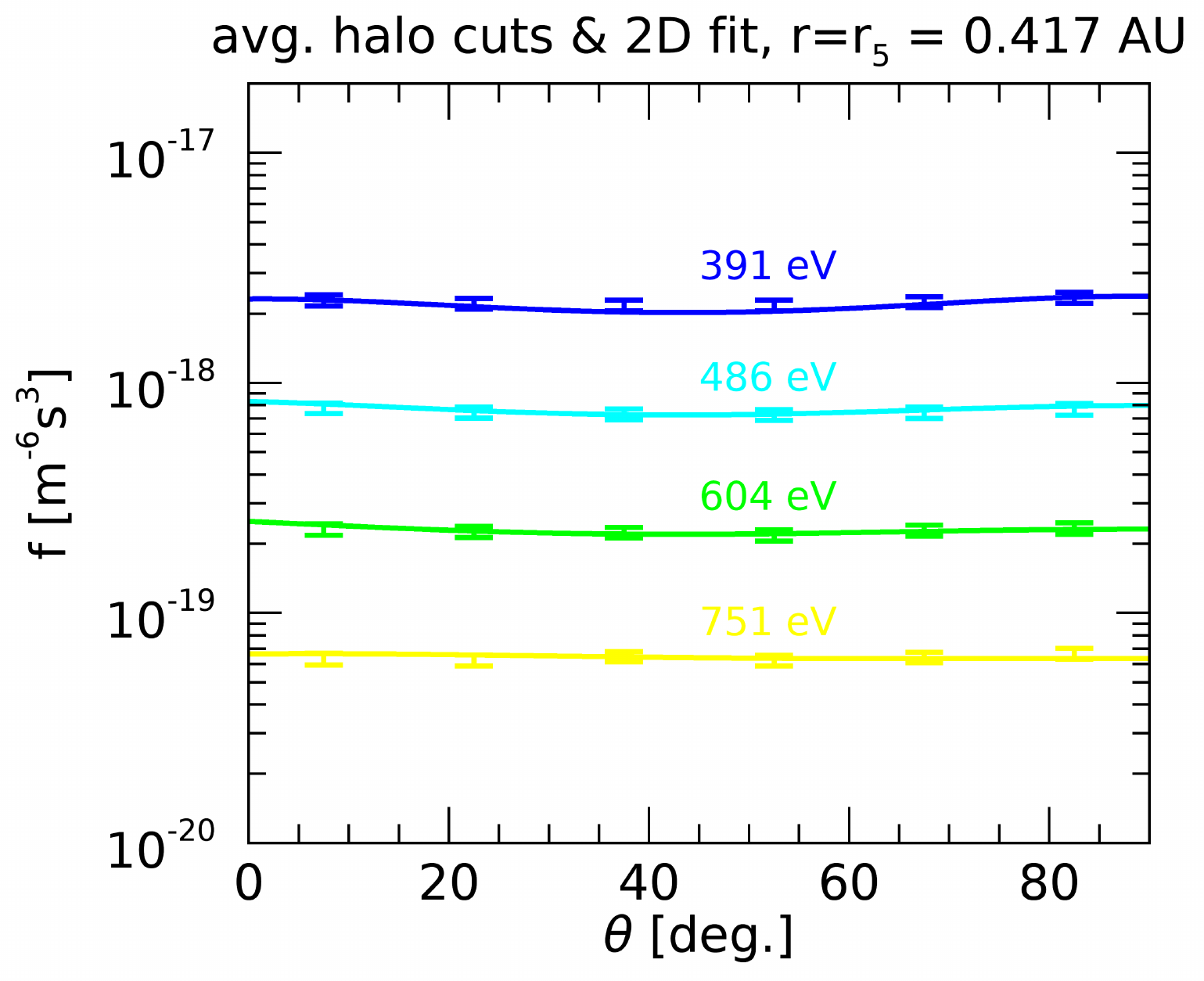}\hspace{1cm}
    \includegraphics[width=.35\textwidth]{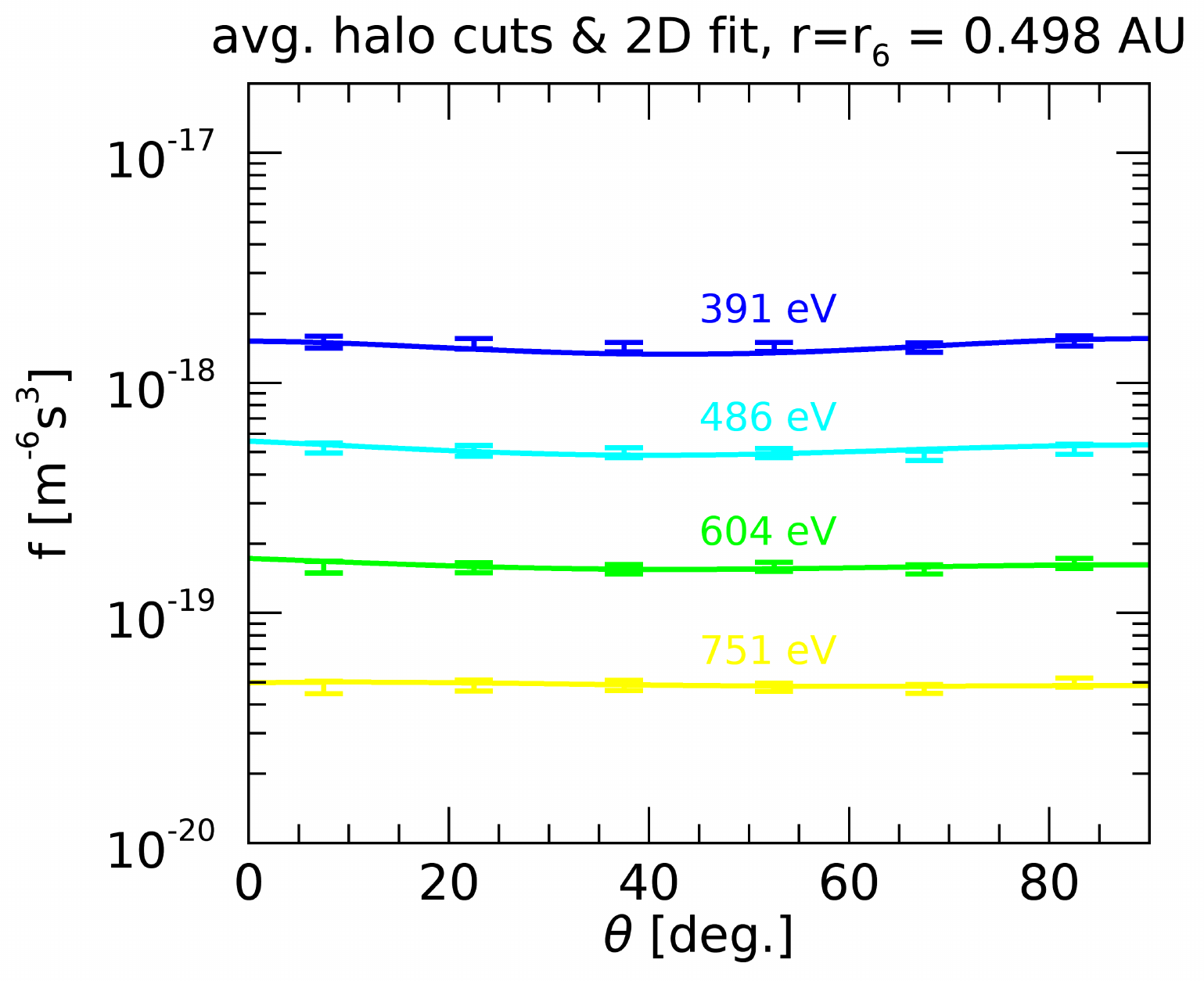}\hfill
    \\[\smallskipamount]
    \includegraphics[width=.35\textwidth]{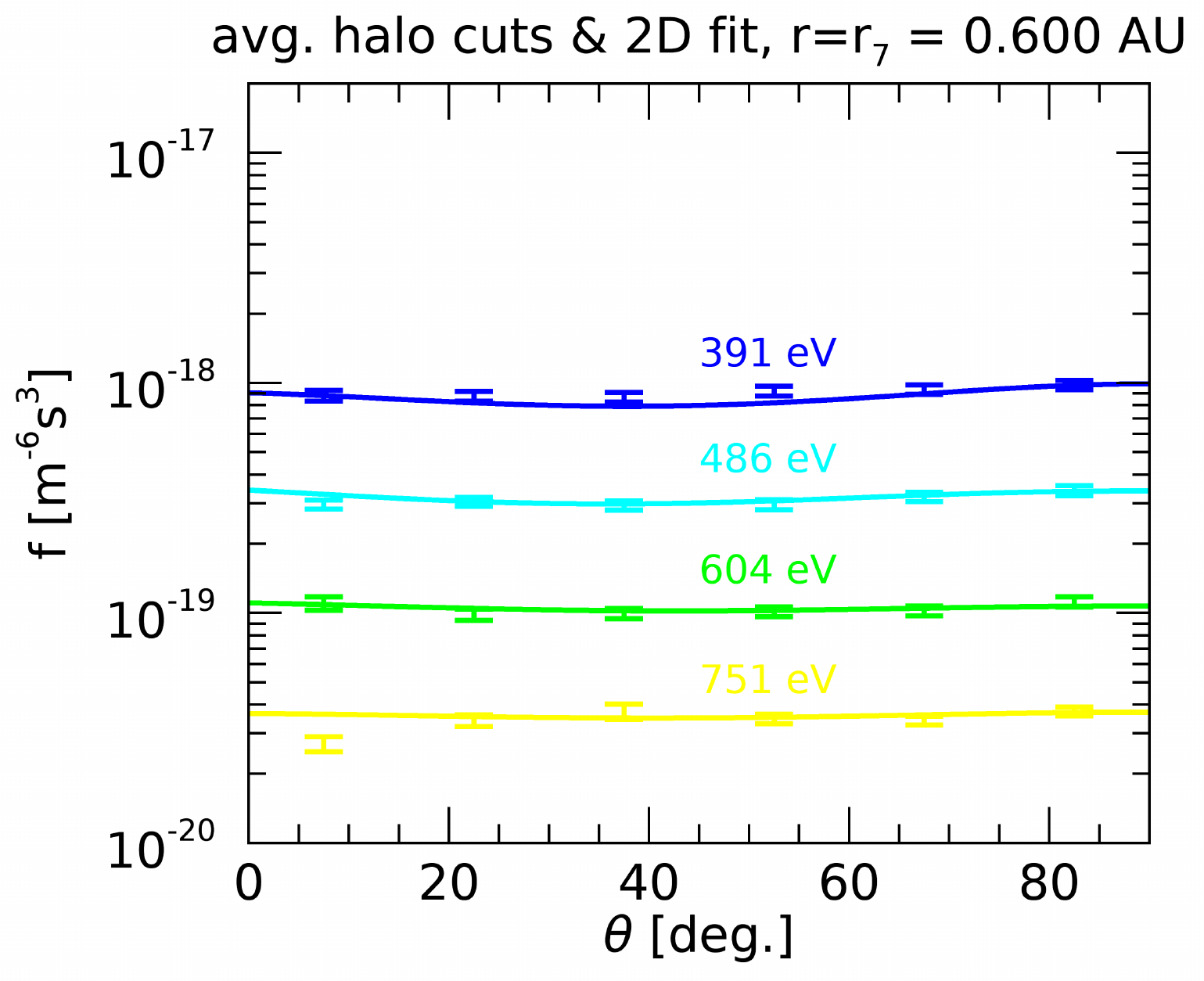}\hspace{1cm}
    \includegraphics[width=.35\textwidth]{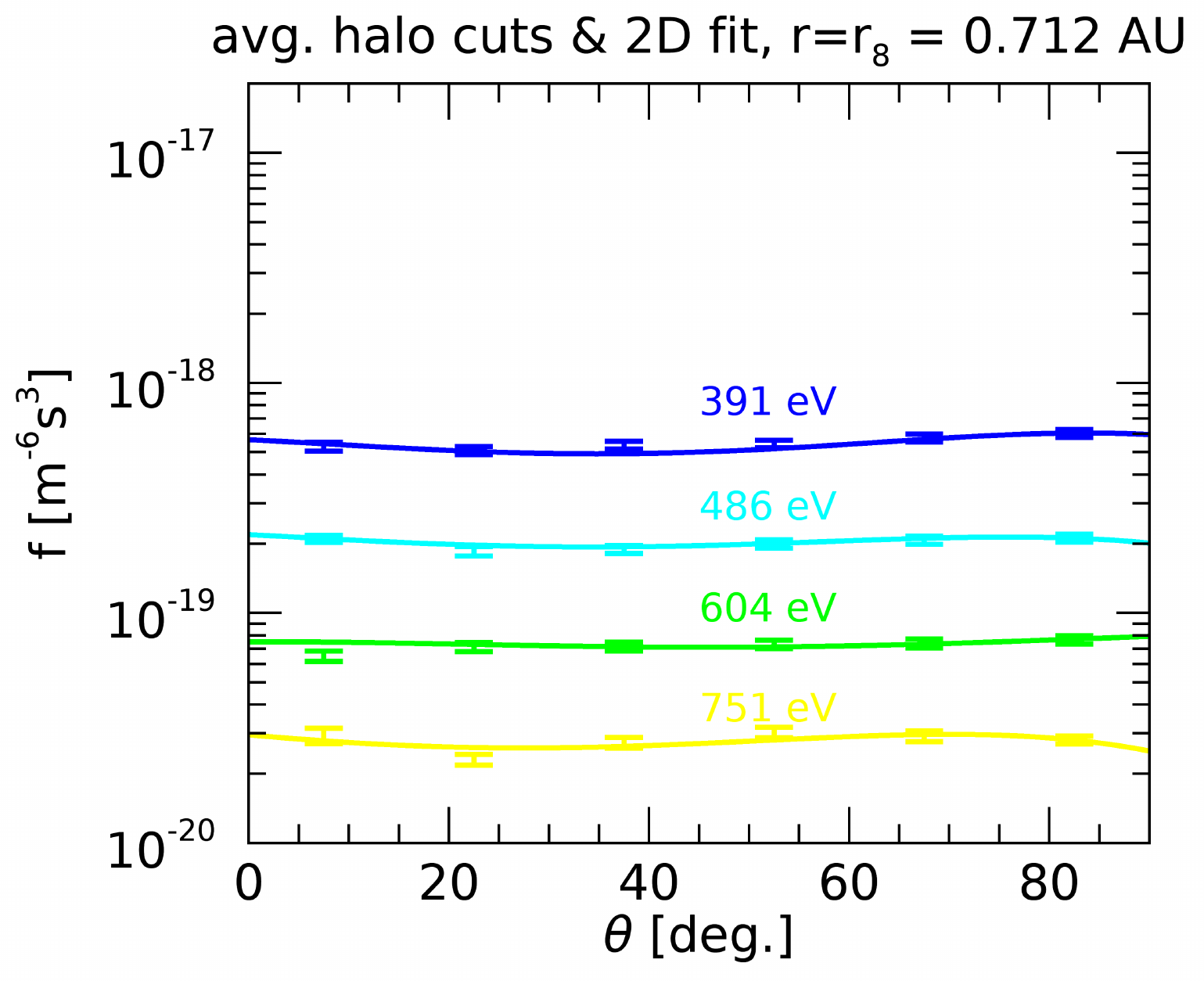}
    \\[\smallskipamount]
    \caption{\label{2d_fit_fig}Halo electron pitch angle distributions ($0^\circ<\theta<90^\circ$, bin width $\Delta \theta$=15$^\circ$) measured by SPAN-E, averaged at $N$=8 distance $r_1$, $r_2$, ...$r_8$. 
    Each panel represents a different distance, and different cuts of constant energy are distinguished by color. We considered PAD data in the energy range [\lowenergy,\highenergy] eV and phase space densities within [2e-20,1e-17]~m$^{-6}$s$^3$. A single non-linear least squares fit is performed to all data shown (fit parameters listed in Tables~ \ref{plasma_param_and_phi_table}, \ref{2d_fit_table}). The fit uses a flexible model equation (\ref{f_fit_K_theta_eq}), a 2D polynomial that is constrained to satisfy Liouville's theorem. The electron energization $e\Delta\phi$ manifests in these plots as a vertical translation in phase space density of a given energy cut (color) between distances (panels). 
    }
\end{figure*}

\noindent \red{Applying the Liouville mapping formula~(\ref{liouville_map_eq}) to the boundary condition (\ref{f_poly_eq}) yields a model equation for $f(r, K,\theta)$, which can be compared directly to the SPAN-E PADs (as they are recorded in terms of energy $K$ and pitch angle $\theta$):}

\begin{equation}\label{f_fit_K_theta_eq}
\begin{split}
            &\ln f(r, K, \theta) = \\
             & \sum_{i=0}^D \sum_{j=0}^D A_{ij} \bigg(\sin^{-1} \sqrt{\frac{ B(r_N)K \sin^2 \theta}{B(r)(K - e\Delta\phi(r,r_N))}} \bigg)^i \Big(K - e\Delta\phi(r,r_N)\Big)^j , \\
\end{split}
\end{equation}

\noindent In the empirical formula~(\ref{f_fit_K_theta_eq}), the terms are given in the following units: [f] = m$^6$s$^{-3}$, [$K$]=eV, [$e\Delta\phi$]=eV, [$\sin^{-1}(x)$]=deg. so that the fit coefficients $A_{ij}$ are unitless.

We fit a single 3D function~(\ref{f_fit_K_theta_eq}) simultaneously to the N distance-averaged SPAN-E PADs, as measured at the distances ${r_1,r_2,..., r_N}$. 
We apply a 4th-degree polynomial (D=4) to match the data. To evaluate the magnetic field $B(r)$ as it appears in eq.~\ref{f_fit_K_theta_eq}, we use PSP's FIELDS fluxgate magnetometer data (averaged by distance bin, see Table~\ref{plasma_param_and_phi_table}). 
There are $(D+1)^2 + N-1$ fit parameters in total: the coefficients $A_{ij}$ and potentials $\phi_k \equiv \Delta \phi(r_1, r_k)$ for $k=2,3,...N$ (note $\phi_1$=0). The latter can be substituted into eq.~\ref{f_fit_K_theta_eq} using the identity:

\begin{equation}
        \Delta \phi(r_k, r_N) = \phi_N - \phi_k,
\end{equation}

\noindent \red{to obtain an equation written in terms of the fit parameters $A_{ij}$, $\phi_k$ that can be matched to the PAD data.}

A nonlinear least-squares fit is performed using the Levenberg-Marquardt algorithm, as implemented in the MPFIT software \citep{markwardt09}. \red{The fit is seeded with initial guesses for the coefficients $A_{ij}$, which are generated by directly fitting the polynomial~(\ref{f_poly_eq}) to the PAD measured the outer distance $r$=$r_N$. Through trial and error we found that an optimal goodness-of-fit could be achieved by considering only energies in the range [\lowenergy,\highenergy] eV. This removes $<$20\% of our halo PAD data and has minimal impact ($\lesssim$10\%) on the fit parameters.} The final fit parameters $\phi_k$ and $A_{ij}$ are listed in Tables~\ref{plasma_param_and_phi_table},\ref{2d_fit_table} respectively. \red{As can be seen in Table~\ref{plasma_param_and_phi_table}, the 1D and 2D methods yield similar estimates of the potential $\Delta \phi$.}

The SPAN-E PAD data are plotted in Figure~\ref{2d_fit_fig} and compared with our fit results.  In each panel of the plot we show energy cuts of the halo PAD (0$^\circ$$<$$\theta$$<$90$^\circ$) averaged at a different distance. \red{The fit to eq.~(\ref{f_fit_K_theta_eq}), shown with solid lines in the plot, accurately describes all data shown in the 8 panels. 
We calculate a reduced chi-squared value of $\chi^2 / \nu=\chisqdof$ ($\nu=$\dof~degrees of freedom, p-value \red{0.16}), which is statistically significant. It is especially noteworthy that our fit successfully matched the averaged PADs even though they have very small error bars. As our model function (\ref{f_fit_K_theta_eq}) is constrained to comply with Liouville's theorem, we may infer that Liouville's theorem is consistent with the halo PAD data.}

\red{To further investigate the significance of our results, we checked whether the collisionless model~(\ref{f_fit_K_theta_eq}) could represent the strahl (which is superficially similar to the halo) equally well. This amounted to repeating our 2D fit method, but redefining the pitch angle: ${\theta\rightarrow (180^\circ-\theta)}$. This way, the fit domain $\theta\in[0^\circ,90^\circ]$ corresponded to the anti-sunward half of the eVDF. The converged fit (not shown) clearly did not match the strahl data, and resulted in an unacceptable reduced chi-squared value: \red{$\chi^2/\nu$=20.1 ($\nu$=171} degrees of freedom, p-value $\lll$1). Of course, it is well-known that scattering (Coulomb collisions and/or wave-particle interactions) is required to explain the strahl angular width \cite[e.g., ][]{lemonsfeldman83}, so it is expected that the model function~(\ref{f_fit_K_theta_eq}) cannot explain the radial variation of the strahl. But this illustrates that Liouville's theorem is highly restrictive---it cannot be satisfied by arbitrary eVDF data.}



\subsection{Proton Acceleration}

The large-scale ambipolar potential is known to accelerate the solar wind proton flow. In the approximation that the potential in the ecliptic is cylindrically symmetric and the protons flow radially outward, a change in potential corresponds directly to a change in radial flow speed. If the ambipolar potential changes with distance as we have inferred from the halo eVDFs (Sections~\ref{1d_sec},~\ref{2d_sec}), then the proton flow energy should change by the same magnitude (up to a small correction due to gravity).

To test this hypothesis, we examine the measurements of the proton flow speed made by SWEAP's SPC Faraday cup. We use the Level~3 velocity data, where for each SPC ion distribution a single 1-dimensional Maxwellian fit was applied in the inertial RTN frame. From this fit we can extract the radial ("R") component of the proton velocity $v_p$. We then compute the kinetic energy of the proton flow: $K_p = m_p v_p^2 / 2$, where $m_p$ is the proton mass. The measurements of $K_p$ are averaged into \nhour-hour intervals and subsequently averaged by distance, as was done with the electron PADs. This gives a proton energy measurement $K_p(r_k)$ at each distance $r_k$ (see Table~\ref{plasma_param_and_phi_table}).

\red{

\begin{figure}
\includegraphics[width=1\linewidth]{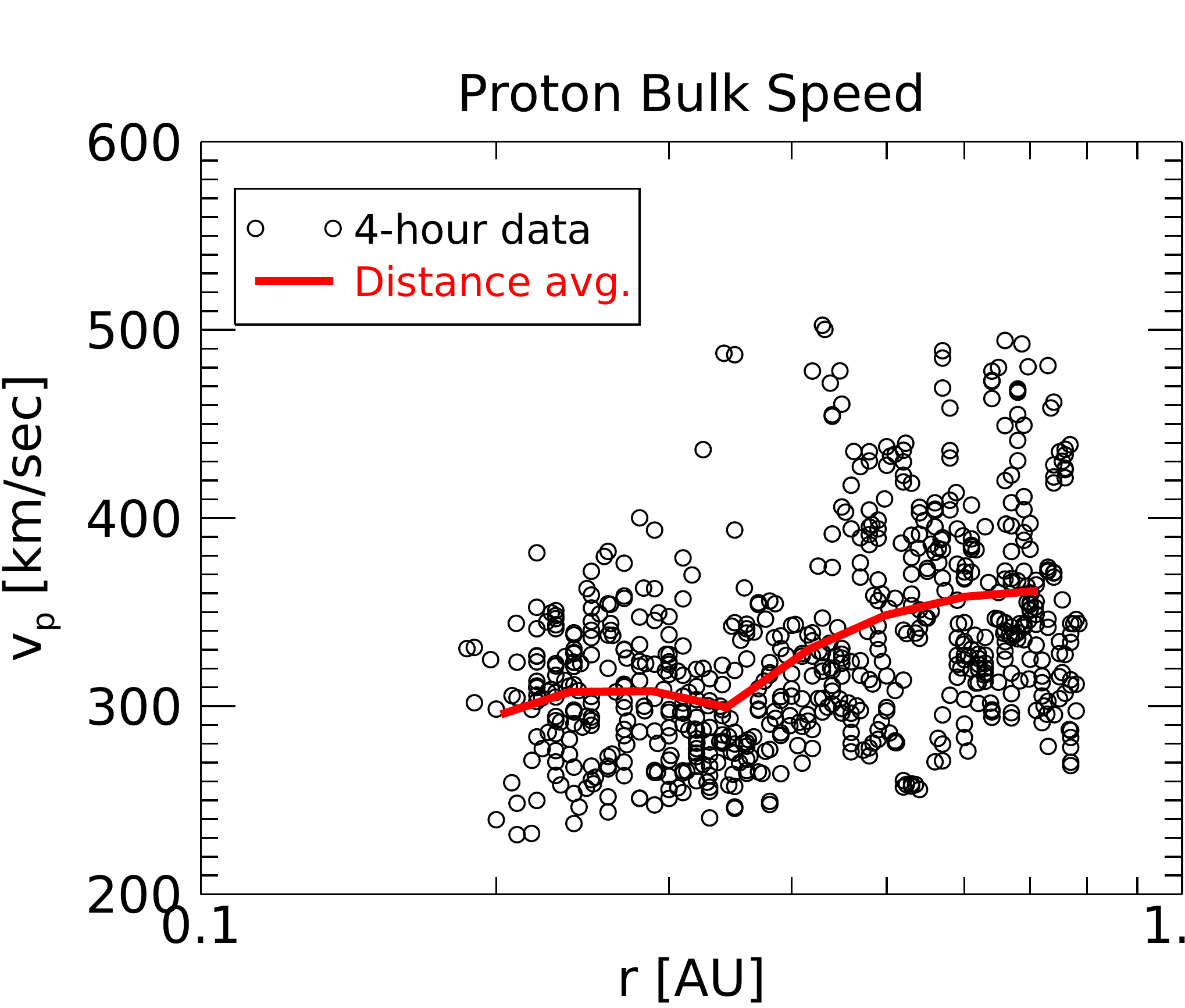}
    \\[\smallskipamount]
\includegraphics[width=1\linewidth]{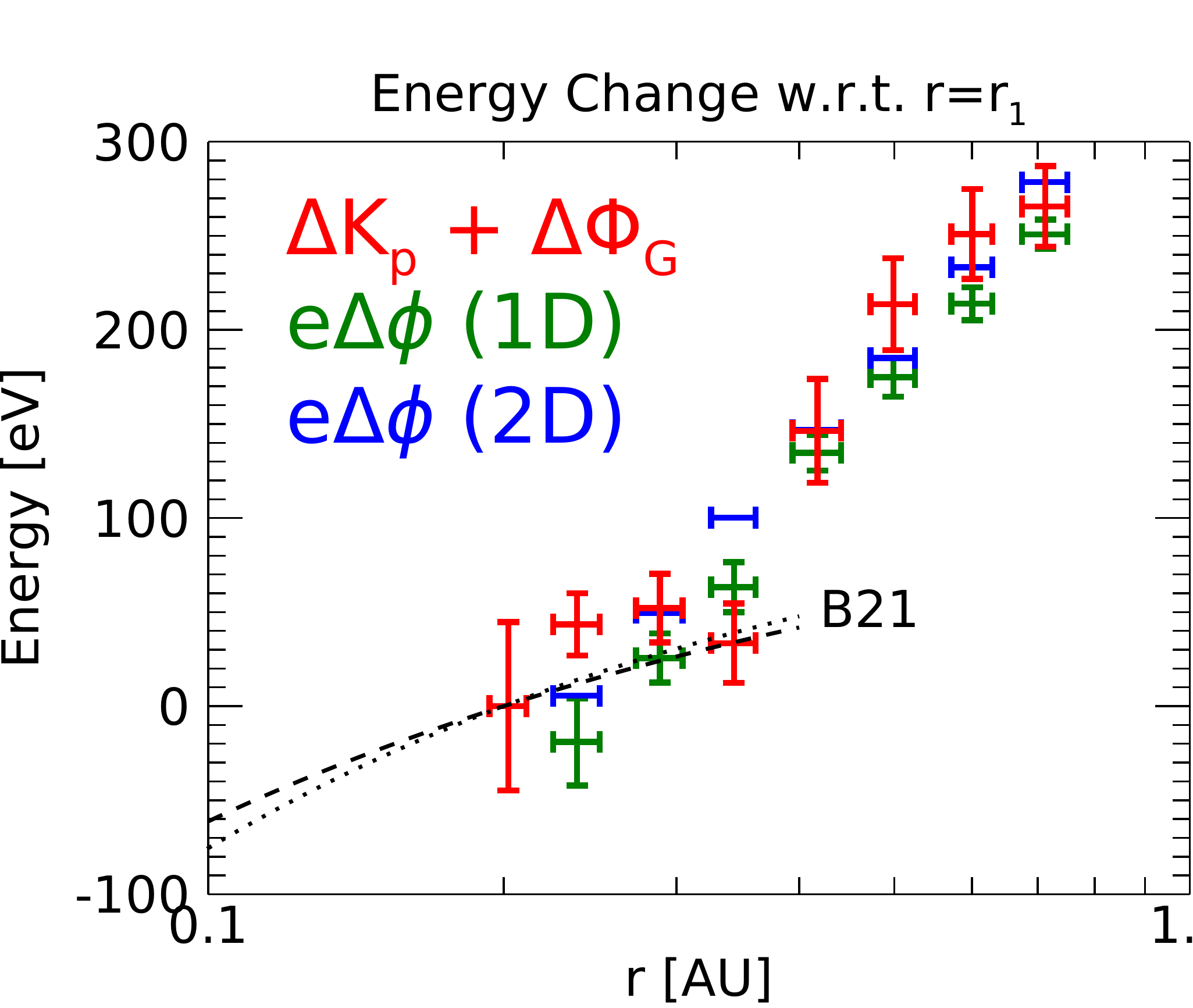}

\caption{\label{phi_and_penergy_fig} 
{\bf Top:} The \nhour-hour averages of the radial proton bulk speed $v_p$, as measured by the SPC Faraday Cup, are shown as circles. The red line shows the average $v_p$ binned by distance (at $r_k$). The data show a steady increasing trend, except the dip around $r_4\approx$0.34~AU, which we treat as an outlier. {\bf Bottom:} \red{The change in electric potential $\Delta \phi(r_1,r_k)$ inferred in two different ways: 1) from the change in proton kinetic energy $\Delta K_p$, correcting for gravity (red, eq.~\ref{delta_phi_proton_eq}), and 2) from the electron halo via the 1D (green) and 2D (blue) methods}.  The two independent estimations of the potential agree within the error bars. The change in potential is given with respect to the minimum distance $r_1=$\rone~AU. 
Two lines are plotted 0.1-0.4~AU, representing the variation of the potential $e\Delta\phi(r_1,r)$ derived from B21, for comparison (see text).  } 
\end{figure}

Because the protons (here treated as a mono-energetic beam) conserve their total (kinetic+potential) energy, the change in electric potential $\Delta \phi$ is given by the formula:

\begin{equation}\label{delta_phi_proton_eq}
    e \Delta \phi(r_1, r_k) = \Delta K_p(r_k) + \Delta \Phi_G(r_k),
\end{equation}

\noindent where we have defined $\Delta K_p(r_k)$ and $\Delta \Phi_G(r_k)$ as the changes in kinetic and gravitational potential energy, relative to the innermost distance $r_1$:

\begin{equation}\label{delta_kp_eq}
    \Delta K_p(r_k) \equiv K_p(r_k)-K_p(r_1),
\end{equation}

\begin{equation}\label{delta_phig_eq}
\Delta \Phi_G(r_k) \equiv G m_S m_p\Big(\frac{1}{r_1} - \frac{1}{r_k}\Big).
\end{equation}

\noindent In eq.~(\ref{delta_phig_eq}), $G$ is Newton's constant and $m_S$ is the solar mass. When the gravitational potential is small, we see from~(\ref{delta_phi_proton_eq}) that a change in electric potential directly causes the same amount of proton acceleration: $\Delta K_p \approx \Delta \phi$. Note though that for protons the gravitational energy cannot be completely neglected, as $\Delta \Phi_G \approx$40~eV across the distances 0.18-0.79~AU.

}

\red{In Fig.~\ref{phi_and_penergy_fig}, we compare the change in potential $e\Delta \phi(r_1, r_k)$ as it is calculated independently in two ways: 1) from the proton kinetic energy $K_p$ (corrected by gravity, eq.~\ref{delta_phi_proton_eq}) and 2) from the  halo eVDFs (1D and 2D methods). The two estimates agree well at all distances 0.18-0.79~AU, implying a change in potential $\Delta \phi \sim$270~eV over the entire interval. This agreement implies that the electric potential inferred from the halo eVDFs fully explains proton acceleration over this distance interval. In rough numbers, the change in electric potential $\sim$270~eV causes the halo electrons to lose this same amount of kinetic energy, while the total (electric+gravitational) potential causes the protons to gain $\sim$230~eV in bulk flow energy.} In terms of velocity, on average the protons start with radial speeds $v_p$$\approx$290~km/sec at 0.18~AU and increase to $v_p$$\approx$360~km/sec at 0.79~AU.

\begin{table*}
\centering
\begin{tabular}{ | c | c | c | c || c | c | }
\hline
$r$ range $[AU]$ & $r_k$ [AU] & B($r_k$) [nT] & $K_p(r_k)$ [eV] & 1D Method: $\Delta\phi(r_1,r_k)$ [V] & 2D Fit: $\phi_k \equiv \Delta\phi(r_1,r_k)$ [V] \\ 
\hline
$[0.18,0.21]$ & $r_1$ = 0.202 &54.3 $\pm$ 5.5 & 463 $\pm$ 31&-0.00 $\pm$ 0 & 0\\
$[0.21,0.26]$ & $r_2$ = 0.237 &37.9 $\pm$ 0.9 & 499 $\pm$ 11&-18.9 $\pm$ 23.2 & 5.5457373\\
$[0.26,0.31]$ & $r_3$ = 0.288 &26.0 $\pm$ 0.8 & 501 $\pm$ 12&25.56 $\pm$ 13.0 & 49.565330\\
$[0.31,0.37]$ & $r_4$ = 0.343 &19.3 $\pm$ 0.7 & 478 $\pm$ 17&63.26 $\pm$ 13.2 & 100.22172\\
$[0.37,0.45]$ & $r_5$ = 0.417 &12.4 $\pm$ 0.3 & 586 $\pm$ 24&134.6 $\pm$ 9.48 & 146.75335\\
$[0.45,0.54]$ & $r_6$ = 0.498 &10.8 $\pm$ 0.3 & 649 $\pm$ 21&174.8 $\pm$ 10.4 & 185.08152\\
$[0.54,0.65]$ & $r_7$ = 0.600 &8.20 $\pm$ 0.2 & 684 $\pm$ 20&213.9 $\pm$ 8.74 & 233.34487\\
$[0.65,0.79]$ & $r_8$ = 0.712 &6.69 $\pm$ 0.1 & 696 $\pm$ 18&250.7 $\pm$ 7.80 & 278.53871\\
\hline
\end{tabular}
\caption{Summary of physical parameters.---\red{(clean up values)} We divide the range of distances in our data set [0.18,0.79] AU into logarithmically spaced bins, labeled in the column ``r range [AU]''. We average over the available within each distance bin data to get a nominal position $r_k$, magnetic field $B(r_k)$, and proton bulk flow energy $K_p(r_k)$---errors of these quantities are calculated as the standard deviation of the mean. Data are only considered into these averages at times when electron PADs are also available. As described in the text, for each distance $r_k$ the potential difference $\Delta \phi(r_1,r_k)=\phi(r_1)-\phi(r_k)$ is computed via the 1D and 2D method. In the 2D method, the potentials $\phi_k$ are fit parameters used in our model (\ref{f_fit_K_theta_eq}). }\label{plasma_param_and_phi_table}
\end{table*}

\begin{table*}
\begin{tabular}{ | c |}
\hline
2D Fit Parameters $A_{ij}$ \\
\hline
\end{tabular}
\\
\centering
\begin{tabular}{ | l | l | l | l | l |}
\hline
$A_{00}$= -36.939956 &  $A_{01}$= -0.018878924 &  $A_{02}$= 2.7295349e-05 &  $A_{03}$= -4.1321067e-08 &  $A_{04}$= 2.5958121e-11 
\\   
$A_{10}$= -0.12204923 &  $A_{11}$= 0.0013464076 &  $A_{12}$= -5.1785826e-06 &  $A_{13}$= 8.1686931e-09 &  $A_{14}$= -4.5181963e-12 
\\   
$A_{20}$= 0.0021237555 &  $A_{21}$= -3.0146726e-05 &  $A_{22}$= 1.2182674e-07 &  $A_{23}$= -1.9396898e-10 &  $A_{24}$= 1.0711158e-13 
\\   
$A_{30}$= 1.4938766e-05 &  $A_{31}$= -2.5832770e-10 &  $A_{32}$= -1.4007037e-10 &  $A_{33}$= 2.5217069e-13 &  $A_{34}$= -1.2619921e-16 
\\   
$A_{40}$= -3.3124890e-07 &  $A_{41}$= 2.7152059e-09 &  $A_{42}$= -9.5596288e-12 &  $A_{43}$= 1.4803501e-14 &  $A_{44}$= -8.2264039e-18 
\\   
\hline
\end{tabular}
\caption{ Fit coefficients $A_{ij}$ used in the 2D method, eq.~(\ref{f_fit_K_theta_eq}). Formal errors are very small, so are not reported. }\label{2d_fit_table}
\end{table*}

\section{Discussion}\label{discussion_sec}


It is appropriate to compare our measurements of the potential with the results of B21. In that work, the authors report that the large-scale potential varies as $\phi(r) = \Phi_0 (r/R_S)^{\alpha_\Phi}$ in the interval 0.1$\lesssim$$r$$\lesssim$0.4 AU. They report two pairs of fit parameters: \{$\Phi_0$=1556.64~V, $\alpha_\Phi$=-0.66\} and \{$\Phi_0$=1043.88~V, $\alpha_\Phi$=-0.55\}.  From these profiles we calculate the change in potential $\Delta\phi(r_1, r)$, which we plot for comparison as dashed/dotted lines in Figure~\ref{phi_and_penergy_fig}. \red{ The B21 profiles agree roughly with our measurements of e$\Delta\phi$ in the interval where both techniques were applied (0.2-0.4 AU). The error bars, however, are comparable to the signal for these small energies ($\lesssim 50$ eV).} 
We note that in our measurements $\Delta \phi$ changes by $>$100 volts in the interval [0.4,0.8]~AU, while extrapolation of B21 yields only $\Delta \phi$$\approx$30--40 volts over this same interval. An increase of hundreds of volts between 0.2 and 0.8 AU is not at all unreasonable in the slow wind. Such an increase may indeed be expected for the protons to accelerate from their modest $\sim$300~km/sec speeds observed at 0.2 AU to their typical $\sim$1~AU values~$\sim$400km/sec \citep{mcgregor11}. 

Our results displayed in Fig.~\ref{phi_and_penergy_fig} indicate that the acceleration of the slow solar wind is almost entirely due to the ambipolar electric field. \red{Gravity is found to have a significant impact as well, and should increase in importance near the Sun \citep{lamy03}}.
In the same sense, gravitational forces diminish rapidly with distance and may be entirely neglected in the outer heliosphere. As may be seen from Ulysses measurements \citep[e.g.][]{stverak09}, the halo itself does evolve slowly in the outer heliosphere. This evolution could be due in part or in whole to the radial variation of the ambipolar potential. 

The Liouville mapping technique accurately describes the halo eVDF. This implies that halo electrons are not affected by diffusive wave-particle interactions in the inner heliosphere. This is in itself an important result, as many theories of halo generation presuppose some local wave mechanism, that for example could scatter the strahl population into the halo. However, the absence of halo and/or strahl diffusion is consistent with the recent measurements of \citet{cattell22, jeong22} which respectively show that whistler and FM/W waves do not scatter the strahl near the Sun.  Our results do not preclude the occasional action of instabilities that derive their free energy from the halo particles, which have been observed e.g., at 1 AU \citep{tong19}. \red{But to zeroth order we may infer that the average sunward halo eVDFs observed by PSP are not locally affected by these instabilities.}

If the sunward-moving halo electrons observed 0.18-0.79 AU are not produced locally by wave-particle diffusion, then they must have originated from some larger heliocentric distance. Such a mechanism of halo generation has not been deeply explored in the present body of research. However, as suggested in \citet{horaites19}, if a sunward-moving suprathermal population is formed in the outer heliosphere, the process of magnetic mirroring should cause it to appear nearly isotropic in the inner heliosphere.

\section{Summary and Conclusions}\label{conclusions_sec}

Using PSP data, we have shown that sunward-moving halo electrons evolve in accordance to Liouville's theorem in the inner heliosphere. This provides a very simple description of the halo dynamics. \red{The potential $\phi(r)$ is the only quantity not measured in situ that is needed in order to map the halo eVDF from one location to another.  This allowed us to apply the Liouville mapping technique to estimate $\phi$.}  We have independently measured the energy of the proton bulk flow, and found that the inferred potential has exactly the energy (within $\sim$10\%) required to accelerate the slow solar wind.

\red{Our measurements of the ambipolar potential are similar to those provided by B21, at least at heliocentric distances where both techniques were applied. Extrapolating the power laws $\phi(r)\sim r^{\alpha_\Phi}$ provided in B21 to distances $r\gtrsim$0.4 AU underestimates the change of potential $\Delta \phi$ compared to our analysis.}
As our approach is built on Liouville's theorem, it rests on firmer theoretical ground than the core deficit approach in its current state of development. From an observational standpoint, the two methods are complementary, \red{and may even inform each other.} It is not feasible to apply our method at distances $r\lesssim$0.2~AU with available data, because of the practical concern of instrument noise at energies $>$500~eV when PSP's mechanical attenuator is engaged. On the other hand, the subtle measurements of the core deficit are reported to be infeasible at distances $\gtrsim$0.4 AU.
\red{A more detailed comparison of these methods is beyond the scope of this paper.}

We have shown that significant solar wind acceleration occurs between 0.2--0.8~AU.  \red{Based on a wind speed of $\sim$400 km/sec at 1~AU during solar minimum \citep{mcgregor11}, we may expect the potential to decrease by another $\sim$100--200~eV outside of PSP's aphelion, $r_N\sim$0.8~AU. This means the halo energy shift caused by the large-scale potential may still be observable at distances $\gtrsim$$r_N$.}
Unfortunately as the protons reach their asymptotic speed the acceleration will likely become even more difficult to discern in the variable solar wind data.  It is worth noting as well that at distances $r$$>$30 AU the solar wind actually decelerates, reportedly due to the pickup of interstellar material \citet{elliott19}.

\red{We have assumed that the sunward-moving halo electrons evolve collisionlessly in the inner heliosphere, without experiencing wave-particle interactions or any other effect that could invalidate Liouville's theorem as it is applied here.} This might seem like a great leap. \red{Wave-particle interactions are often invoked as a mechanism that could plausibly account for both the halo's isotropy and its evolution with distance.} However, our model also meets these requirements.  The isotropy can be explained in terms of the mirror force, which broadens the PAD and also reflects the sunward-moving particles into an anti-sunward distribution. \red{The radial evolution is directly explained, as our fit function has been matched to the halo eVDF at all observed distances (Fig.~\ref{2d_fit_fig}). The Liouville mapping's effectiveness suggests that the sunward halo did not experience local scattering during the PSP measurements.} We additionally infer that if the strahl electrons undergo wave-particle interactions in the inner heliosphere, these interactions do not significantly influence the sunward-moving electrons.

The present work applies a simple collisionless model to explain how the average halo eVDFs observed by PSP evolve in the inner heliosphere. \red{Our results also support the basic premise of exospheric theories, that the proton flow speed is dictated by the large-scale potentials.} However, if this model is correct, we require an explanation for how the seed population of energetic, sunward-moving electrons is formed in the outer heliosphere (or beyond). This highly motivating question can be addressed in future research.




\section*{Acknowledgements}

We acknowledge the NASA Parker Solar Probe Mission and the SWEAP team (led by J.~Kasper) and FIELDS team (led by S.~D.~Bale) for use of data. We thank S.~D.~Bale and D.~Larson for feedback on the particle measurements.  The work of SB was partly supported by NSF Grant PHY-2010098, by NASA Grant NASA 80NSSC18K0646, and by the Wisconsin Plasma Physics Laboratory (US Department of Energy Grant DE-SC0018266)

\section*{Data Availability}

All PSP data were obtained from the Coordinated Data Analysis Web (CDAWeb) service \url{https://cdaweb.gsfc.nasa.gov/}. The MPFIT software for IDL can be downloaded at \url{http://cow.physics.wisc.edu/~craigm/idl/idl.html}. The SIR/CIR event list can be found at \url{https://sppgway.jhuapl.edu/Event_List}. 
The HELIO4CAST ICMECAT catalog is published on the data sharing platform figshare  \url{https://doi.org/10.6084/m9.figshare.6356420} and at \url{https://helioforecast.space/icmecat}. Here the version 2.1, updated on 2021 December 7 was used (this is version 11 on figshare).





\bibliographystyle{mnras}
\bibliography{paper_refs} 

\bsp	
\label{lastpage}
\end{document}